\documentclass[astrosymb]{aastex701}

\shorttitle{Extremely low mass ratio contact binaries}
\shortauthors{Lai et al.}
\submitjournal{ApJ}

\usepackage{amsmath}
\usepackage[T1]{fontenc}
\begin{document}
	\title{Extremely Low Mass Ratio Contact Binaries. III. Photometric and Spectroscopic Investigations of Eleven Systems}

	\author[orcid=0009-0005-9831-9152,gname=Youmo,sname=Lai]{Youmo Lai}
	\email{202400830133@mail.sdu.edu.cn}
	\affiliation{Shandong Key Laboratory of Optical Astronomy and Solar-Terrestrial Environment, School of Space Science and Technology, Institute of Space Sciences, Shandong University, Weihai, Shandong 264209, People's Republic of China}
	
	\author[orcid=0000-0003-3590-335X,gname=Kai,sname=Li]{Kai Li}
	\email{kaili@sdu.edu.cn}
	\correspondingauthor{Kai Li}
	\affiliation{Shandong Key Laboratory of Optical Astronomy and Solar-Terrestrial Environment, School of Space Science and Technology, Institute of Space Sciences, Shandong University, Weihai, Shandong 264209, People's Republic of China}
	\affiliation{International Centre of Supernovae (ICESUN), Yunnan Key Laboratory, Kunming 650216, China}
		
	\author[orcid=0000-0003-1263-808X,gname=Raul,sname={Michel Murillo}]{Raul Michel Murillo}
	\email{rmm@astro.unam.mx}
	\affiliation{Instituto de Astronom\'ia, Universidad Nacional Aut\'onoma de M\'exico, Apdo. Postal 106, Ensenada 22800, Baja California, M\'exico}
		
	\author[orcid=0000-0003-4957-485X,gname=Difu,sname=Guo]{Difu Guo}
	\email{difu@sdu.edu.cn}
	\affiliation{Shandong Key Laboratory of Optical Astronomy and Solar-Terrestrial Environment, School of Space Science and Technology, Institute of Space Sciences, Shandong University, Weihai, Shandong 264209, People's Republic of China}

	\author[orcid=0009-0005-0485-418X,gname=Liheng,sname=Wang]{Liheng Wang}
	\email{243786559@qq.com}
	\affiliation{Shandong Key Laboratory of Optical Astronomy and Solar-Terrestrial Environment, School of Space Science and Technology, Institute of Space Sciences, Shandong University, Weihai, Shandong 264209, People's Republic of China}

	\author[orcid=0009-0009-6364-0391,gname=Xiang,sname=Gao]{Xiang Gao}
	\email{229391256@qq.com}
	\affiliation{Shandong Key Laboratory of Optical Astronomy and Solar-Terrestrial Environment, School of Space Science and Technology, Institute of Space Sciences, Shandong University, Weihai, Shandong 264209, People's Republic of China}

	\author[orcid=0009-0008-9589-6044,gname=Han-ning,sname=Zhang]{Han-ning Zhang}
	\email{202400830122@mail.sdu.edu.cn}
	\affiliation{Shandong Key Laboratory of Optical Astronomy and Solar-Terrestrial Environment, School of Space Science and Technology, Institute of Space Sciences, Shandong University, Weihai, Shandong 264209, People's Republic of China}

	\author[orcid=0000-0002-7292-3109,gname=Xing,sname=Gao]{Xing Gao}
	\email{34052688@qq.com}
	\affiliation{Xinjiang Astronomical Observatory, 150 Science 1-Street, Urumqi 830011, People's Republic of China}

	\author[orcid=0000-0003-3162-3350,gname=Guo-You,sname=Sun]{Guo-You Sun}
	\email{36723571@qq.com}
	\affiliation{Xingming Observatory, Urumqi, Xinjiang, People's Republic of China}

	\begin{abstract}

		We present photometric and spectroscopic investigations of 11 totally eclipsing contact binaries with mass ratios below 0.15, classifying them as extremely low mass ratio contact binaries. The ground-based multiband light curves were modeled with PHysics Of Eclipsing BinariEs to derive the photometric parameters of these systems. The TESS light curves of J011323 and J002747 show pronounced asymmetries and continuous variations. Markov Chain Monte Carlo modeling suggests that these variations are related to longitudinal migration of spots on the primary components. Spectral subtraction of the LAMOST spectra reveals excess chromospheric emission in six systems, while no detectable excess emission is found in the other five. The O$-$C analysis indicates secular period increases in six systems and secular decreases in five systems. We estimated their orbital angular momenta and performed a dynamical analysis. {Within our adopted semi-analytical framework without energy transfer, the formation of these observed systems requires additional angular momentum loss beyond saturated magnetic braking and gravitational radiation, but this requirement could be modified by different assumptions on mass transfer or radius evolution.}

	\end{abstract}
	\keywords{\uat{Eclipsing binary stars}{444} --- \uat{Contact binary stars}{297} --- \uat{Fundamental parameters of stars}{555} --- \uat{Multiple star evolution}{2153}}

\section{Introduction}
\label{sec:introduction}

Contact binaries are systems in which both components fill their Roche lobes
and share a common envelope \citep{Lucy1968}. They are also known
as W UMa-type binaries and are usually composed of F-, G-, or K-type stars
\citep{Rucinski1993}. Their contact configuration and tidal distortion produce
continuously varying light curves, with primary and secondary eclipses that
often have comparable depths \citep{Kuiper1941,Lucy1967,MatteiSaladyga1999}.
Contact binaries are commonly
classified as A- or W-subtype systems: the more massive component is hotter in
A-subtype binaries and cooler in W-subtype binaries \citep{Binnendijk1970}.

W UMa-type binaries often show unequal out-of-eclipse maxima, known as the
O'Connell effect \citep{OConnell1951}. The asymmetry may arise from cool spots
associated with magnetic activity \citep{Mullan1975,Wilsey2009}, hot spots produced by mass
accretion \citep{Shaw1994,Fabry2025a}, circumstellar material \citep{LiuYang2003}, or flow
asymmetry driven by the coriolis effect \citep{ZhouLeung1990}. This effect can remain stable in some
systems but vary irregularly or quasi-periodically in others
\citep{Balaji2015,Yilmaz2023}. The time-dependent asymmetry may therefore be
interpreted as spot migration caused by solar-like differential rotation
\citep{Tran2013}.

The mass ratio is central to both the evolutionary state and dynamical stability
of a contact binary. Extremely low mass ratio contact binaries (ELMRCBs) are
defined by $q<0.15$ \citep{Li2022ELMRCBI} and are regarded as possible
progenitors of blue stragglers or FK Com-type stars
\citep{Eggleton2012,YangQian2015}. Under Darwin's instability hypothesis, a
contact binary is expected to merge once the spin angular momentum exceeds
one-third of the orbital angular momentum \citep{Hut1980}. For unevolved
main-sequence contact binaries, \citet{Rasio1995} derived an instability
threshold of $q_{\rm inst}\simeq0.09$. Subsequent work showed that the critical
mass ratio depends on stellar structure and the adopted gyration radii; for
example, \citet{LiZhang2006b} obtained $q_{\rm inst}\simeq0.071$--0.078.
\citet{Arbutina2009} considered rotation-induced central condensation with a
rotating polytrope model and lowered the limit to 0.070--0.074.
\citet{Wadhwa2021} related the primary's gyration radius $k_1$ to its mass
$M_1$, yielding a global minimum of $\approx0.042$--0.044.
\citet{arbutina2024critical} further discussed the use of the Darwin instability
criterion to identify potential merger candidates.
\citet{Wadhwa2024} showed that the instability mass ratio decreases
with decreasing metallicity. \citet{Zhang2024} derived a global minimum of
$q_{\min}\approx0.038$--0.041. Using asteroseismic constraints on internal rotation and gyration radii, \citet{Zhang2026} obtained a critical mass ratio of $0.042$--0.044.

Observational and statistical studies have also challenged the classical
limits. \citet{YangQian2015} inferred a possible lower limit of
$q_{\rm min}\simeq0.044$ from statistical extrapolation. \citet{Pesta2023}
showed that the statistical lower limit depends on subtype and orbital period,
with $q_{\rm min}\simeq0.087$ for late-type systems with $P>0.3$ days,
$q_{\rm min}\simeq0.246$ for shorter-period late-type systems, and
$q_{\rm min}\simeq0.030$ for early-type systems. \citet{ELMRCBII} further estimated an empirical
cut-off mass ratio of $q_{\rm min}\simeq0.021$ from the relation between
fill-out factor and mass ratio. Several confirmed systems now
approach or fall below the quoted thresholds, including V857 Her ($q\simeq0.065$; \citealt{Qian2006}), M4 V53
($q\simeq0.078$; \citealt{Li2017}), V1187 Her ($q\simeq0.044$; \citealt{Caton2019}), VSX J082700.8+462850
($q\simeq0.055$; \citealt{Li2021}), TYC 4002-2628-1
($q\simeq0.048$; \citealt{Guo2022}), TYC 3801-1529-1
($q\simeq0.036$, \citealt{Li2024a}; $q\simeq0.024$, \citealt{Poro2026}), and the
$q\simeq0.027$ system reported by \citet{Guo2025}. These objects show that the
low mass ratio limit remains an open problem requiring both larger
observational samples and refined stability calculations.

For totally eclipsing contact binaries, photometric light-curve modeling can
provide reliable mass-ratio estimates even without radial-velocity data
\citep{Rucinski2001,Pribulla2003,TerrellWilson2005}. Motivated by the
empirical connection between low photometric amplitudes and low mass ratios
\citep{Rucinski2001,Pesta2023}, we selected 11 low-amplitude, totally
eclipsing contact binaries from the All-Sky Automated Survey for Supernovae
(ASAS-SN; \citealt{Shappee2014,Jayasinghe2018}). Their basic information is listed
in Table~\ref{tab:basic_information}. J011323 has been studied by
\citet{Popov2022}; the other 10 systems are analyzed for the first time.
This work is the third paper in our ELMRCB series, following Paper I
\citep{Li2022ELMRCBI} and Paper II \citep{ELMRCBII}. We combine multiband
photometry, LAMOST spectroscopy, orbital-period analysis, and evolutionary
diagnostics to characterize the 11 systems.

\begin{deluxetable*}{clllcccccc}[ht!]
	\tabletypesize{\scriptsize}
	\tablewidth{0pt}
	\tablecaption{Basic information of the 11 targets \label{tab:basic_information}}
	\tablehead{
		\colhead{Target} & \colhead{Hereafter} & \colhead{Other name} & \colhead{Period} & \colhead{Min. (HJD)} & \colhead{Mean $V$mag} & \colhead{Amplitude} & \colhead{RUWE} & \colhead{E(B-V)}
	}
	\startdata
	ASASSN-V J002747.72+371506.8 & J002747 & CSS\_J002747.6+371507 & 0.3211590 & 2459885.35021 & 14.85 & 0.32 & 1.472 & 0.080$\pm$0.007 \\
	ASASSN-V J003357.06+362917.8 & J003357 & CSS\_J003357.0+362917 & 0.3246479 & 2459889.87687 & 15.13 & 0.32 & 1.057 & 0.040$\pm$0.007 \\
	ASASSN-V J004106.41+455050.6 & J004106 & WISEJ004106.4+455050 & 0.3503785 & 2459883.25222 & 14.43 & 0.33 & 1.042 & 0.080$\pm$0.005 \\
	ASASSN-V J011323.69+374319.2 & J011323 & GSC 02800-01387 & 0.3026253 & 2460613.22122 & 14.18 & 0.34 & 0.992 & 0.040$\pm$0.005 \\
	ASASSN-V J024047.31+543946.3 & J024047 & 2MASS J02404729+5439463 & 0.3426494 & 2460635.27897 & 14.91 & 0.18 & 1.083 & 0.350$\pm$0.009 \\
	ASASSN-V J025641.85+363505.3 & J025641 & CSS\_J025641.7+363503 & 0.3221738 & 2460648.27361 & 14.70 & 0.30 & 1.092 & 0.100$\pm$0.010 \\
	ASASSN-V J030224.38+300429.4 & DK Ari & DK Ari & 0.3087263 & 2460668.18249 & 13.99 & 0.28 & 1.030 & 0.150$\pm$0.005 \\
	ASASSN-V J040106.16+182220.9 & J040106 & CSS\_J040106.2+182221 & 0.2852765 & 2459894.76484 & 15.31 & 0.28 & 1.244 & 0.360$\pm$0.012 \\
	ASASSN-V J085750.84+400057.8 & J085750 & CSSJ085750.7+400058 & 0.4242533 & 2459634.17606 & 12.98 & 0.34 & 7.847 & 0.050$\pm$0.012 \\
	ASASSN-V J091721.13+482908.4 & J091721 & VSX J091721.0+482908 & 0.3428096 & 2459631.14256 & 13.56 & 0.37 & 0.952 & 0.018$\pm$0.007 \\
	ASASSN-V J232802.69+233657.6 & J232802 & ASAS J232803+2336.9 & 0.4120807 & 2459850.37847 & 12.48 & 0.30 & 1.028 & 0.070$\pm$0.007 \\
	\enddata
	\tablecomments{RUWE is the Renormalised Unit Weight Error from \cite{Gaia2016,Gaia2021}; the Gaia EDR3 data are available from \citet{10.5270/esa-1ugzkg7}. E(B-V) and its uncertainty are from the Bayestar2019 3D dust map \citep{Green2019}.}
\end{deluxetable*}

	\section{Observations}
	\label{sec:Obs}
	\subsection{Photometric Observations}

From 2022 to 2024, we obtained multiband photometry of the 11 targets with
four ground-based facilities. The Weihai Observatory 1.0 m telescope of
Shandong University (WHOT; \citealt{Hu2014}) is equipped with a PIXIS 2048B CCD
camera, with a pixel scale of about $0\farcs35$ pixel$^{-1}$ and a field of
view of about $12\arcmin \times 12\arcmin$. The
85 cm telescope at the Xinglong Station of National Astronomical Observatories
(XL85) uses an Andor DZ936 CCD camera, with a pixel scale of about
$0\farcs94$ pixel$^{-1}$ and a field of view of about
$32\arcmin \times 32\arcmin$. The 60 cm Ningbo Bureau of Education and
Xinjiang Observatory Telescope (NEXT) uses a back-illuminated FLI PL23042 CCD
camera and has a field of view of about $22\arcmin \times 22\arcmin$. The
84 cm telescope at the Observatorio Astron\'omico Nacional San Pedro M\'artir
(SPM84) is equipped with an E2V back-illuminated CCD detector, with a
pixel scale of $0\farcs26$ pixel$^{-1}$ and a field of view of about
$9\arcmin \times 9\arcmin$. Standard Johnson--Cousins or Sloan filters were used
depending on the instrument. The observing log is summarized in
Table~\ref{tab:obslog}.

	All CCD frames were reduced with IRAF using standard bias subtraction,
flat-field correction, and aperture photometry. Stable field stars were selected
as comparison (C) and check (CH) stars. We derived differential
magnitudes for both the target--comparison pair (V-C) and the
comparison--check pair (C-CH), and adopted the standard deviation of C-CH as the photometric uncertainty.

	We also retrieved public survey photometry from several archives: calibrated
full-frame images from the Transiting Exoplanet Survey Satellite (TESS;
\citealt{Ricker2014,10.17909/0cp4-2j79}),
the Zwicky Transient Facility (ZTF; \citealt{Bellm2019,10.26131/irsa598}),
the Super Wide Angle Search for Planets (SuperWASP; \citealt{Butters2010,10.26133/nea9}),
the Catalina Real-time Transient Survey (CRTS; \citealt{Drake2009,10.26093/cds/vizier.16960870}),
and ASAS-SN \citep{Shappee2014,Jayasinghe2018}. The survey data were used
mainly to measure additional eclipse minima for the O$-$C analysis. The
continuous TESS light curves were also used to analyze short-timescale
light-curve variability.

	\begin{deluxetable*}{lllllll}[ht!]
		\tabletypesize{\scriptsize}
		\tablewidth{0pt}
		\tablecaption{Ground-based photometric observing log \label{tab:obslog}}
		\tablehead{
			\colhead{Target} & \colhead{Telescope} & \colhead{Observation(s) Date} & \colhead{Exposure time(s)} & \colhead{Comparison star} & \colhead{Check star} & \colhead{Mean errors (mag)}
		}
		\startdata
		J002747 & NEXT & 2022 Oct 21, Nov 01 & g100 r90 & \shortstack[l]{2MASS\\00274131+3714539} & \shortstack[l]{2MASS\\00280649+3712309} & g0.007 r0.008 \\
		J003357 & SPM84 & 2022 Oct 05 & B120 V90 R60 & \shortstack[l]{2MASS\\00342208+3631136} & \shortstack[l]{2MASS\\00342194+3631544} & B0.008 V0.006 R0.006 \\
		J004106 & NEXT & 2022 Oct 30, Nov 01, 02 & g100 r90 & \shortstack[l]{2MASS\\00410838+4548079} & \shortstack[l]{2MASS\\00410539+4545419} & g0.005 r0.004 \\
		J011323 & XL85 & 2024 Oct 29 & R70 I60 & \shortstack[l]{2MASS\\01132100+3744501} & \shortstack[l]{2MASS\\01133106+3744169} & R0.004 I0.005 \\
		J024047 & XL85 & 2024 Nov 20 & R100 I90 & \shortstack[l]{2MASS\\02404687+5438048} & \shortstack[l]{2MASS\\02401430+5438408} & R0.004 I0.004 \\
		J025641 & XL85 & 2024 Dec 03 & R100 I90 & \shortstack[l]{2MASS\\02563523+3635312} & \shortstack[l]{2MASS\\02562530+3636568} & R0.004 I0.005 \\
		DK Ari & WHOT & 2024 Dec 23 & V100 R60 I50 & \shortstack[l]{2MASS\\03023016+3005289} & \shortstack[l]{2MASS\\03022596+3003574} & V0.013 R0.008 I0.007 \\
		J040106 & SPM84 & 2022 Oct 10 & V90 R60 & \shortstack[l]{2MASS\\04010267+1825345} & \shortstack[l]{2MASS\\04011084+1825164} & V0.008 R0.006 \\
		J085750 & NEXT & 2022 Feb 20, 21, 23 & g35 r30 i40 & \shortstack[l]{2MASS\\08575556+4002006} & \shortstack[l]{2MASS\\08582205+3959180} & g0.005 r0.004 i0.004 \\
		J091721 & WHOT & 2022 Feb 20 & V100 R50 I45 & \shortstack[l]{2MASS\\09173675+4832454} & \shortstack[l]{2MASS\\09174353+4826324} & V0.011 R0.011 I0.012 \\
		J232802 & NEXT & 2022 Sep 23, 27 & g25 r22 i30 & \shortstack[l]{2MASS\\23280352+2332313} & \shortstack[l]{2MASS\\23281528+2340019} & g0.008 r0.008 i0.009 \\
		\enddata
	\end{deluxetable*}

	\subsection{Spectroscopic Observations}

	The Large Sky Area Multi-Object Fiber Spectroscopic Telescope (LAMOST) is a
4 m reflecting Schmidt telescope with a $5^\circ$ field of view and 4000
fibers on its focal plane \citep{Cui2012}. We cross-matched the 11 targets with
LAMOST Data Release 12 using a radius of $3\arcsec$ and retrieved all available
low-resolution spectra ($R\sim1800$, 3700--9000\,\AA). Several targets have
multiple low-resolution spectra, but none has a matching medium-resolution
spectrum in the archive. For each spectrum, we adopted the DR12 atmospheric
parameters, including effective temperature ($T_{\mathrm{eff}}$), metallicity
([Fe/H]), the $g$-band signal-to-noise ratio ($SNR_g$), radial velocity (RV),
and surface gravity ($\log g$), as listed in Table~\ref{tab:spre}.

	\begin{deluxetable*}{lcccrrrc}[ht!]
		\tabletypesize{\scriptsize}
		\tablewidth{0pt}
		\tablecaption{LAMOST parameters of the targets \label{tab:spre}}
		\tablehead{
			\colhead{Target} & \colhead{Observing Date} & \colhead{Type} & \colhead{$SNR_g$} & \colhead{$T_{\mathrm{eff}}$ (K)} & \colhead{[Fe/H]} & \colhead{RV (km s$^{-1}$)} & \colhead{$\log g$}
		}
		\startdata
		J002747 & 2014/11/22 & G3 & 62 & $5717 \pm 29$ & $0.043 \pm 0.024$ & $-31.35 \pm 3.53$ & $4.173 \pm 0.031$ \\
		& 2012/10/31 & G3 & 32 & $5712 \pm 41$ & $0.023 \pm 0.041$ & $-21.23 \pm 3.20$ & $4.127 \pm 0.034$ \\
		J003357 & 2020/11/16 & F2 & 11 & $5970 \pm 115$ & $-0.394 \pm 0.125$ & $-14.07 \pm 2.82$ & $4.126 \pm 0.030$ \\
		& 2022/11/29 & F7 & 27 & $6066 \pm 67$ & $-0.207 \pm 0.069$ & $-14.73 \pm 2.28$ & $4.047 \pm 0.024$ \\
		J004106 & 2014/10/18 & F6 & 94 & $6082 \pm 28$ & $-0.154 \pm 0.023$ & $2.17 \pm 4.67$ & $4.174 \pm 0.061$ \\
		& 2012/10/06 & F6 & 27 & $6185 \pm 49$ & $-0.084 \pm 0.050$ & $-21.00 \pm 4.36$ & $4.246 \pm 0.060$ \\
		J011323 & 2015/10/29 & G3 & 75 & $5770 \pm 21$ & $0.092 \pm 0.018$ & $0.70 \pm 3.37$ & $4.052 \pm 0.035$ \\
		& 2015/10/29 & F9 & 48 & $5660 \pm 17$ & $0.030 \pm 0.015$ & $-2.91 \pm 4.07$ & $4.129 \pm 0.043$ \\
		J024047 & 2017/01/24 & F3 & 54 & $6247 \pm 42$ & $-0.559 \pm 0.036$ & $-72.66 \pm 3.44$ & $4.152 \pm 0.036$ \\
		J025641 & 2013/10/10 & A8 & 73 & $6849 \pm 28$ & $-0.507 \pm 0.024$ & $5.94 \pm 5.02$ & $4.138 \pm 0.107$ \\
		DK Ari & 2014/11/14 & A6 & 66 & $6814 \pm 36$ & $-0.783 \pm 0.030$ & $-103.77 \pm 7.64$ & $4.232 \pm 0.114$ \\
		& 2014/11/19 & F0 & 27 & $6698 \pm 75$ & $-0.746 \pm 0.077$ & $-111.17 \pm 4.05$ & $4.407 \pm 0.038$ \\
		J040106 & 2014/12/17 & F8 & 45 & $5748 \pm 33$ & $-0.652 \pm 0.029$ & $20.35 \pm 3.72$ & $4.185 \pm 0.039$ \\
		& 2014/12/24 & F6 & 34 & $5784 \pm 49$ & $-0.666 \pm 0.047$ & $5.56 \pm 3.81$ & $4.026 \pm 0.040$ \\
		& 2014/09/26 & G0 & 23 & $5775 \pm 50$ & $-0.619 \pm 0.053$ & $20.09 \pm 4.81$ & $4.073 \pm 0.057$ \\
		J085750 & 2023/11/16 & F5 & 65 & $6398 \pm 28$ & $-0.004 \pm 0.024$ & $-1.16 \pm 5.87$ & $4.063 \pm 0.065$ \\
		& 2013/11/19 & F5 & 127 & $6423 \pm 23$ & $0.016 \pm 0.017$ & $-12.18 \pm 4.61$ & $4.237 \pm 0.079$ \\
		& 2013/11/19 & F5 & 69 & $6411 \pm 24$ & $0.038 \pm 0.020$ & $-9.94 \pm 5.81$ & $4.238 \pm 0.061$ \\
		J091721 & 2014/03/21 & F0 & 69 & $6334 \pm 26$ & $-0.579 \pm 0.022$ & $20.11 \pm 7.58$ & $4.144 \pm 0.192$ \\
		& 2017/02/09 & F0 & 17 & $6447 \pm 65$ & $-0.801 \pm 0.070$ & $36.02 \pm 6.90$ & $4.155 \pm 0.101$ \\
		& 2012/04/03 & F3 & 30 & $6318 \pm 42$ & $-0.525 \pm 0.042$ & $-6.05 \pm 5.65$ & $4.067 \pm 0.060$ \\
		& 2012/04/03 & F3 & 29 & $6350 \pm 41$ & $-0.620 \pm 0.041$ & $13.18 \pm 4.79$ & $4.232 \pm 0.051$ \\
		J232802 & 2012/10/01 & F7 & 38 & $6146 \pm 26$ & $0.087 \pm 0.024$ & $-36.56 \pm 3.89$ & $4.084 \pm 0.039$ \\
		& 2013/10/15 & F7 & 55 & $6148 \pm 30$ & $0.048 \pm 0.026$ & $-35.26 \pm 4.93$ & $4.088 \pm 0.074$ \\
		\enddata
	\end{deluxetable*}

	\section{Light-curve Solutions}
	\label{sec:Sol}

\subsection{Preliminary Parameter Estimation}

All light curves were modeled with PHysics Of Eclipsing BinariEs
(PHOEBE) version 2.4
\citep{PrsaZwitter2005,Prsa2016,Horvat2018,Conroy2020,Jones2020}.
To obtain suitable starting values for the MCMC sampling, we first used
the Contact Binary Light-curve Analyzer (CBLA; \citealt{Li2025a}), a
neural-network tool that rapidly provides PHOEBE-compatible estimates
of the mass ratio $q$, orbital inclination $i$, fill-out factor $f$,
temperature ratio $T_2/T_1$, and spot parameters. Before the parameter
search, the primary temperature was fixed to the mean spectroscopic
temperature from LAMOST. Throughout this paper, $q=M_2/M_1$,
	with star 1 denoting the more massive component.

For each target, the epoch of primary minimum in
	Table~\ref{tab:basic_information} was determined with the
	\citet{KweeWoerden1956} method and adopted as phase zero. After
removing significant outliers, we converted the differential
magnitudes (V-C) to fluxes and normalized each light curve by the mean
flux near phase 0.25.

We first allowed all adjustable model parameters, including
	$l_3$, to vary. We then compared the free-$l_3$ solutions with
	the corresponding $l_3=0$ solutions using the Bayesian information
	criterion (BIC), $\chi^2$, and $R^2$ based on the residuals between
	the observed and modeled light curves. For most
	targets, allowing $l_3$ to vary did not improve the fits. For the
	three targets with lower BIC values in the free-$l_3$ models, the
	inferred third-light fractions were all below 5\%. The Gaia DR3
source catalog shows no nearby source within $10\arcsec$ of any target
except J024047, whose neighboring sources are too faint to contribute
appreciably. Moreover, the available O$-$C data show no
	significant light-travel-time effect attributable to a third body.
	We therefore adopted $l_3=0$ for the subsequent ground-based
	light-curve modeling. This does not exclude the presence of faint
	tertiary companions in these systems, which will require confirmation
	through future high-precision observations.

With $l_3$ fixed at zero, we then tested separate spot
	configurations on the two components. The spot colatitude $\theta$,
	longitude $\lambda$, angular radius $r_s$, and temperature factor
	$T_s$ were allowed to vary. Cool- and hot-spot solutions were
explored over $T_s=[0.6,1.0]$ and $[1.0,1.4]$, respectively, and the
solution with the larger coefficient of determination ($R^2$) was
adopted. Better fits were obtained when the spot was placed on
	star 1.

\subsection{PHOEBE MCMC Sampling}

After obtaining the preliminary solutions, we performed MCMC sampling
with PHOEBE using the \texttt{emcee} package
\citep{ForemanMackey2013}. Following \citet{Lucy1967} and
\citet{Rucinski1973}, the bolometric albedo and gravity-darkening
coefficients of both components were fixed at 0.5 and 0.32, respectively.

\begin{figure*}[ht!]
	\centering
	\includegraphics[width=0.74\textwidth]{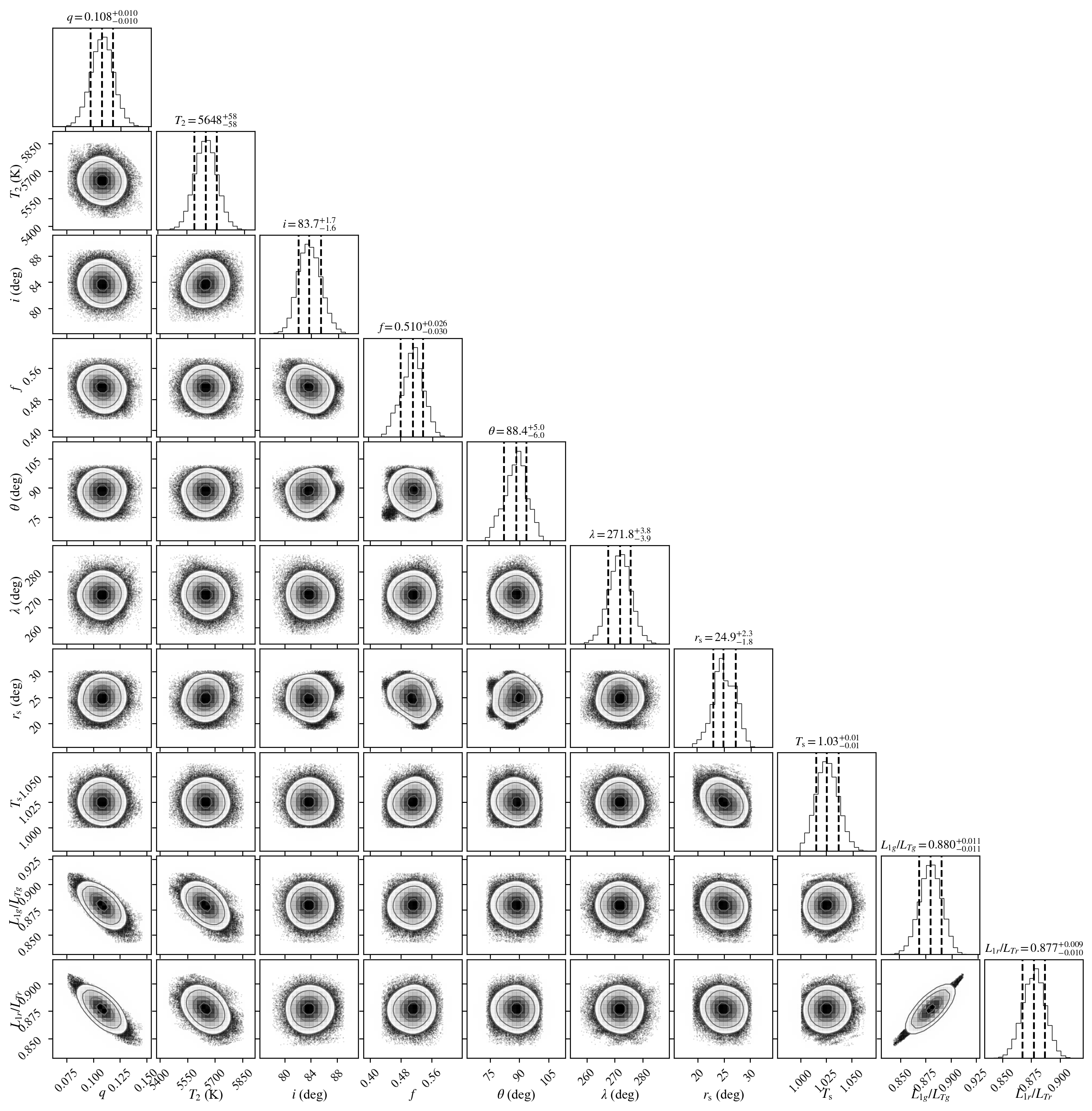}
	\caption{Probability distributions of $q$, $T_2$, $i$, $f$, $L_{1X}/L_{TX}$ and spot parameters determined by the MCMC modeling of J002747.}
	\label{fig:corner_j002747}
	\end{figure*}

Based on the preliminary tests, the third light was fixed at
	$l_3=0$, and the spot was placed on star 1 in the final MCMC
	analysis. Gaussian priors were assigned to the jointly sampled
	parameters $q$, $T_2$, $i$, $f$, $\theta$, $\lambda$, $r_s$, $T_s$,
	and the primary passband luminosity $L_{1X}$ in each observed
	passband $X$. Following \citet{Conroy2020}, convergence was accepted when
$N_{\mathrm{iter}}>10\tau$ for every free parameter, where $\tau$ is
the integrated autocorrelation time. More than 2000 iterations were
performed for each target, and all chains satisfied this criterion.
The posterior distributions of J002747 are shown as a
	representative example in Figure~\ref{fig:corner_j002747}.

	Because the adopted LAMOST temperature represents the composite spectroscopic temperature of the binary, we refined the individual component temperature using the temperature ratio and the following relations (\citealt{Zwitter2003}; \citealt{ChristopoulouPapageorgiou2013}):

\begin{equation}
	\begin{aligned}
		k &= \frac{r_2}{r_1}, \\
		T_1 &= \left[ \frac{(1 + k^2)\,T_{\mathrm{spec}}^4}
		{1 + k^2 \left( \frac{T_{2i}}{T_{1i}} \right)^4} \right]^{1/4}, \\
		T_2 &= T_1 \left( \frac{T_{2i}}{T_{1i}} \right).
	\end{aligned}
	\label{eq:temperature}
\end{equation}

\begin{figure*}[ht!]
	\centering
	\includegraphics[width=\textwidth]{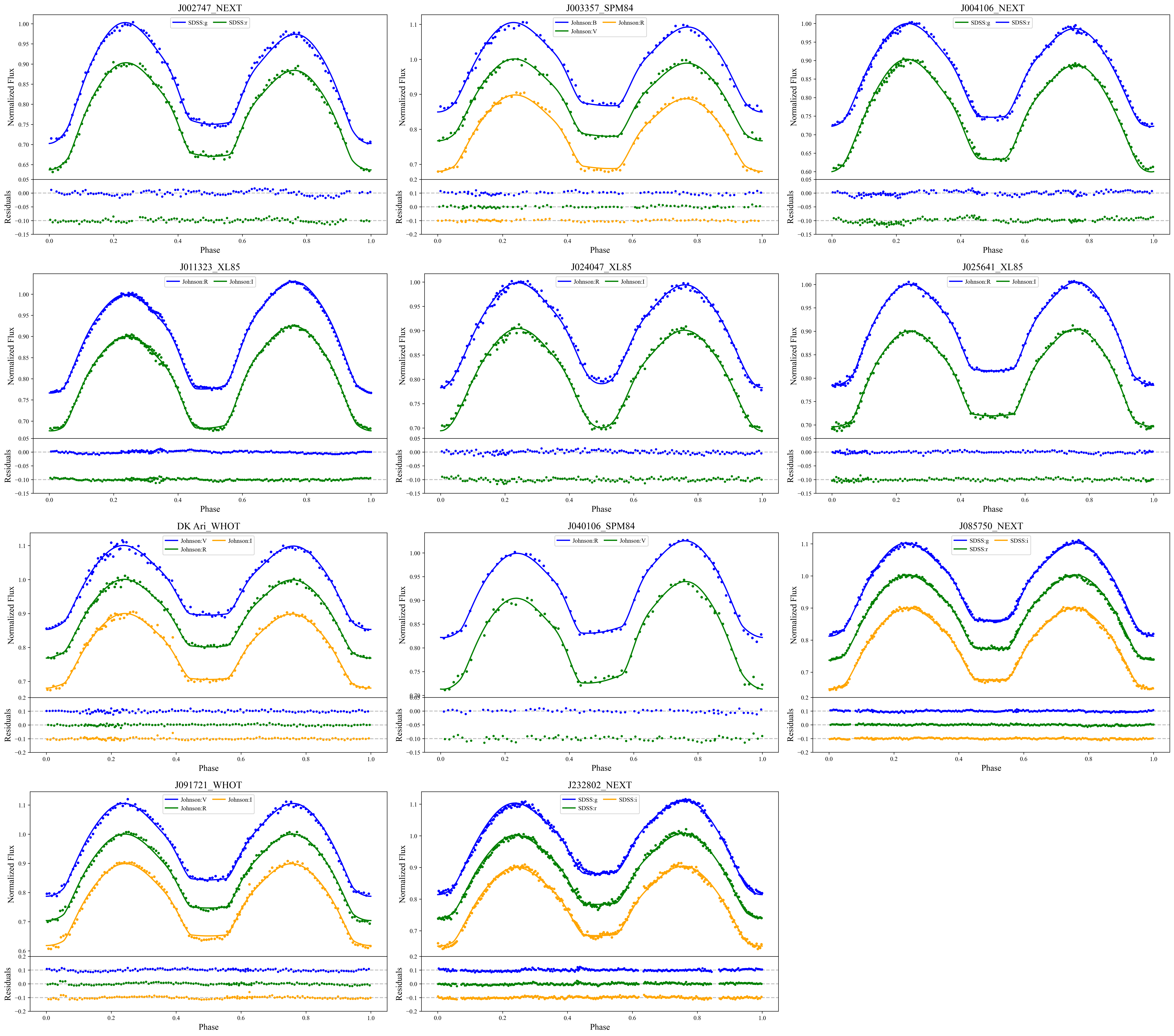}
	\caption{Comparison between the observed and synthetic light curves for the 11 targets. The points represent the observed data, the solid curves represent the fitted models, and the lower panels show the residuals. The data used to create this figure are available in the online version.}
	\label{fig:lc_all}
\end{figure*}

Here $r_1$ and $r_2$ are the PHOEBE volume-equivalent radii,
$T_{\mathrm{spec}}$ is the composite LAMOST temperature, and
$T_{2i}/T_{1i}$ is the converged temperature ratio.
This approximation assumes efficient energy redistribution
	in the common envelope; more general contact-binary energy-transfer
	prescriptions are discussed by \citet{Fabry2023}.
The corrected temperatures and propagated uncertainties are listed in
Table~\ref{tab:results}. The other uncertainties are formal MCMC errors
and represent the internal statistical precision of the fits, so they
should be regarded as lower limits \citep{Conroy2020}. The observed and
synthetic light curves are compared in Figure~\ref{fig:lc_all}. Our
solution for J011323 is broadly consistent with that of
\citet{Popov2022}, although its fill-out factor is slightly lower.
Most light curves are reproduced well; the remaining localized
	residuals, such as those in J232802, may arise from unknown physical
	effects that are not included in the present light-curve model.

\begin{table*}[ht!]
	\centering
	\caption{Photometric solutions for the 11 targets\label{tab:results}}
	\setlength{\tabcolsep}{3pt}
	\scriptsize
	
		\begin{tabular}{lcccccccccc}
			\hline\hline
			Target & $q$ & $T_1$ (K) & $T_2$ (K) & $i$ ($^\circ$) & $f$ & $\theta$ ($^\circ$) & $\lambda$ ($^\circ$) & $r_s$ ($^\circ$) & $T_s$ & $r_1$ \\
			\hline
			J002747 & $0.108_{-0.010}^{+0.010}$ & $5723\pm36$ & $5657\pm68$ & $83.7_{-1.6}^{+1.7}$ & $0.510_{-0.030}^{+0.026}$ & $88.4_{-6.0}^{+5.0}$ & $271.8_{-3.9}^{+3.8}$ & $24.9_{-1.8}^{+2.3}$ & $1.03_{-0.01}^{+0.01}$ & $0.590_{-0.006}^{+0.007}$ \\
			J003357 & $0.089_{-0.010}^{+0.010}$ & $6006\pm91$ & $6107\pm113$ & $77.1_{-1.0}^{+1.0}$ & $0.638_{-0.028}^{+0.028}$ & $61.2_{-4.5}^{+4.5}$ & $117.2_{-5.7}^{+9.2}$ & $7.5_{-0.7}^{+0.7}$ & $0.70_{-0.02}^{+0.02}$ & $0.606_{-0.007}^{+0.008}$ \\
			J004106 & $0.126_{-0.010}^{+0.012}$ & $6123\pm40$ & $6192\pm75$ & $76.5_{-1.0}^{+1.0}$ & $0.692_{-0.026}^{+0.025}$ & $95.8_{-5.7}^{+4.2}$ & $62.8_{-5.0}^{+4.8}$ & $7.3_{-0.7}^{+0.7}$ & $0.67_{-0.03}^{+0.02}$ & $0.585_{-0.005}^{+0.006}$ \\
			J011323 & $0.115_{-0.008}^{+0.013}$ & $5702\pm21$ & $5796\pm65$ & $77.0_{-1.0}^{+1.0}$ & $0.344_{-0.025}^{+0.021}$ & $108.4_{-3.6}^{+3.7}$ & $90.8_{-5.1}^{+4.6}$ & $37.7_{-1.4}^{+1.4}$ & $1.03_{-0.01}^{+0.01}$ & $0.580_{-0.007}^{+0.006}$ \\
			J024047 & $0.113_{-0.011}^{+0.010}$ & $6253\pm43$ & $6207\pm78$ & $69.7_{-0.9}^{+1.0}$ & $0.379_{-0.018}^{+0.019}$ & $114.4_{-4.7}^{+4.7}$ & $70.2_{-5.6}^{+4.4}$ & $10.4_{-0.9}^{+0.9}$ & $0.93_{-0.02}^{+0.02}$ & $0.582_{-0.006}^{+0.007}$ \\
			J025641 & $0.090_{-0.010}^{+0.009}$ & $6868\pm29$ & $6700\pm78$ & $82.8_{-1.3}^{+1.2}$ & $0.587_{-0.030}^{+0.030}$ & $146.1_{-7.3}^{+5.3}$ & $284.1_{-6.0}^{+8.8}$ & $10.5_{-1.4}^{+1.4}$ & $0.78_{-0.03}^{+0.04}$ & $0.604_{-0.007}^{+0.008}$ \\
			DK Ari & $0.092_{-0.011}^{+0.007}$ & $6772\pm59$ & $6638\pm92$ & $79.6_{-1.0}^{+1.0}$ & $0.781_{-0.021}^{+0.022}$ & $103.2_{-6.6}^{+5.2}$ & $24.4_{-7.8}^{+7.4}$ & $4.0_{-1.1}^{+1.2}$ & $0.67_{-0.02}^{+0.02}$ & $0.608_{-0.006}^{+0.007}$ \\
			J040106 & $0.066_{-0.009}^{+0.010}$ & $5748\pm43$ & $5944\pm80$ & $77.5_{-1.2}^{+1.2}$ & $0.508_{-0.028}^{+0.037}$ & $42.2_{-3.2}^{+3.7}$ & $101.1_{-4.5}^{+4.4}$ & $20.2_{-1.4}^{+1.7}$ & $1.08_{-0.02}^{+0.02}$ & $0.622_{-0.009}^{+0.009}$ \\
			J085750 & $0.117_{-0.010}^{+0.012}$ & $6397\pm27$ & $6492\pm82$ & $78.8_{-1.0}^{+1.0}$ & $0.655_{-0.027}^{+0.020}$ & $155.9_{-4.8}^{+2.6}$ & $180.7_{-4.6}^{+4.3}$ & $43.8_{-1.6}^{+1.9}$ & $1.05_{-0.02}^{+0.02}$ & $0.589_{-0.006}^{+0.006}$ \\
			J091721 & $0.137_{-0.011}^{+0.010}$ & $6360\pm36$ & $6234\pm74$ & $86.0_{-1.2}^{+1.4}$ & $0.791_{-0.018}^{+0.024}$ & $91.5_{-4.5}^{+6.2}$ & $3.2_{-2.2}^{+3.2}$ & $4.9_{-1.4}^{+1.2}$ & $0.65_{-0.02}^{+0.03}$ & $0.584_{-0.005}^{+0.005}$ \\
			J232802 & $0.108_{-0.010}^{+0.008}$ & $6147\pm30$ & $6144\pm79$ & $77.7_{-1.0}^{+1.0}$ & $0.795_{-0.029}^{+0.030}$ & $24.8_{-3.1}^{+5.6}$ & $161.5_{-5.2}^{+4.9}$ & $9.9_{-1.3}^{+1.3}$ & $1.29_{-0.03}^{+0.03}$ & $0.599_{-0.006}^{+0.006}$ \\
			\hline
		\end{tabular}
		
		\vspace{1ex}
		
		\begin{tabular}{lcccccccc}
			\hline\hline
			Target & $r_2$ & $L_{1g}/L_{Tg}$ & $L_{1r}/L_{Tr}$ & $L_{1i}/L_{Ti}$ & $L_{1B}/L_{TB}$ & $L_{1V}/L_{TV}$ & $L_{1R}/L_{TR}$ & $L_{1I}/L_{TI}$ \\
			\hline
			J002747 & $0.229_{-0.007}^{+0.006}$ & $0.880_{-0.011}^{+0.011}$ & $0.877_{-0.010}^{+0.010}$ & -- & -- & -- & -- & -- \\
			J003357 & $0.220_{-0.008}^{+0.008}$ & -- & -- & -- & $0.876_{-0.013}^{+0.013}$ & $0.880_{-0.011}^{+0.012}$ & $0.882_{-0.011}^{+0.012}$ & -- \\
			J004106 & $0.248_{-0.007}^{+0.006}$ & $0.844_{-0.012}^{+0.012}$ & $0.847_{-0.011}^{+0.011}$ & -- & -- & -- & -- & -- \\
			J011323 & $0.227_{-0.006}^{+0.006}$ & -- & -- & -- & -- & -- & $0.863_{-0.010}^{+0.010}$ & $0.865_{-0.009}^{+0.009}$ \\
			J024047 & $0.227_{-0.006}^{+0.006}$ & -- & -- & -- & -- & -- & $0.873_{-0.009}^{+0.009}$ & $0.873_{-0.009}^{+0.009}$ \\
			J025641 & $0.219_{-0.008}^{+0.008}$ & -- & -- & -- & -- & -- & $0.895_{-0.010}^{+0.010}$ & $0.894_{-0.009}^{+0.010}$ \\
			DK Ari & $0.228_{-0.009}^{+0.008}$ & -- & -- & -- & -- & $0.891_{-0.010}^{+0.011}$ & $0.889_{-0.011}^{+0.011}$ & $0.888_{-0.010}^{+0.010}$ \\
			J040106 & $0.196_{-0.009}^{+0.009}$ & -- & -- & -- & -- & $0.898_{-0.011}^{+0.011}$ & $0.901_{-0.010}^{+0.010}$ & -- \\
			J085750 & $0.240_{-0.007}^{+0.007}$ & $0.849_{-0.012}^{+0.013}$ & $0.853_{-0.010}^{+0.011}$ & $0.855_{-0.010}^{+0.011}$ & -- & -- & -- & -- \\
			J091721 & $0.259_{-0.007}^{+0.006}$ & -- & -- & -- & -- & $0.853_{-0.010}^{+0.010}$ & $0.850_{-0.009}^{+0.010}$ & $0.848_{-0.009}^{+0.011}$ \\
			J232802 & $0.240_{-0.007}^{+0.008}$ & $0.865_{-0.012}^{+0.012}$ & $0.865_{-0.010}^{+0.011}$ & $0.866_{-0.010}^{+0.010}$ & -- & -- & -- & -- \\
			\hline
		\end{tabular}
	
	\tablecomments{$T_1$ and $T_2$ are the corrected component temperatures calculated using
		Equation~\ref{eq:temperature}. Their uncertainties are propagated from the
		input temperatures and radii, while the other listed uncertainties are formal
		MCMC errors.}
\end{table*}

\subsection{Temporal Light-curve Variations}

The continuous TESS light curves of J011323 and J002747 show
time-dependent asymmetries that reverse within a single sector. To
trace this evolution, we divided each sector into eight temporal
segments, phase-folded each segment, and binned the resulting light
curves into 300 points. This segmentation preserves the gradual evolution
of the light-curve shape while keeping sufficient phase coverage in each
subsample.

\begin{figure*}[ht!]
	\centering
	\includegraphics[width=\textwidth]{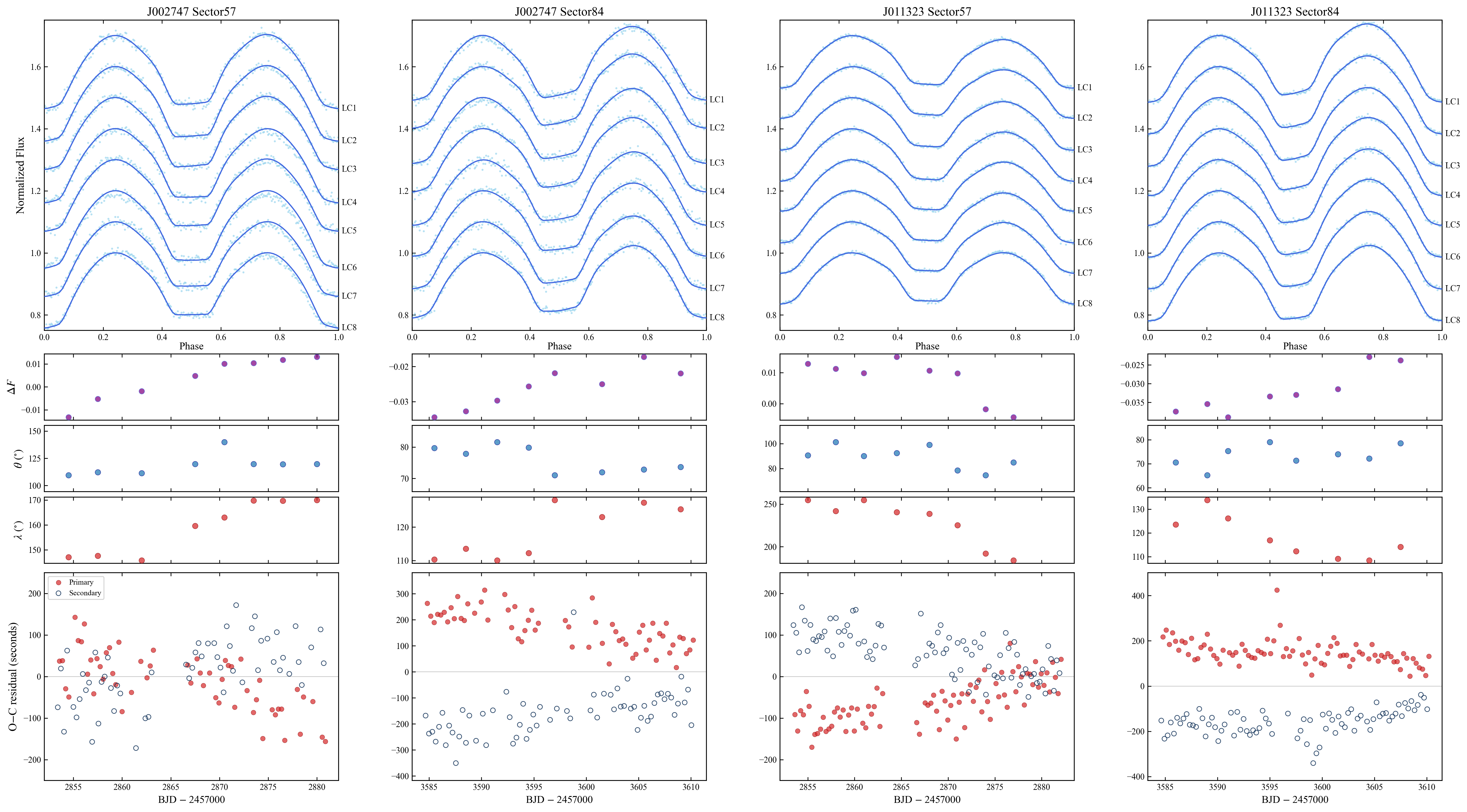}
	\caption{Segmented TESS light-curve fits, O'Connell-effect diagnostics,
		and eclipse-timing residuals for J002747 and J011323. For each target
		and sector, the stacked light curves show the observed data and fitted
		models. The middle panels show the temporal
		variations of $\Delta F$, spot colatitude ($\theta$), and longitude
		($\lambda$). The bottom panels show the O$-$C residuals of primary
		eclipses (red filled circles) and secondary eclipses (blue open circles)
		after removing a common linear trend.}
	\label{fig:oconnell_tess}
	\label{fig:oconnell_oc}
\end{figure*}

We quantified the asymmetry using the flux difference between the two
out-of-eclipse maxima,
$\Delta F=F_{\rm Max\,I}-F_{\rm Max\,II}$, measured from the mean
fluxes near phases 0.25 and 0.75. As shown in
Figure~\ref{fig:oconnell_tess}, $\Delta F$ changes from $-0.0132$ to
$+0.0131$ for J002747 in Sector 57, indicating a transition from
Max II being brighter to Max I being brighter. J011323 shows the
opposite trend, with $\Delta F$ decreasing from $+0.0129$ to
$-0.0043$.

To investigate whether these variations are associated with spot
migration \citep{Kalimeris2002}, we modeled each segment with CBLA,
fixing the primary temperature and allowing the remaining parameters
to vary. In Sector 57, the
spot longitude changed from approximately $255^\circ$ to $184^\circ$
for J011323 and from $147^\circ$ to $170^\circ$ for J002747, whereas
the other fitted parameters showed no comparable monotonic trends.
This systematic longitude evolution suggests spot migration on the
more massive component.

We also examined whether the evolving asymmetry affected the measured
eclipse times. After removing the linear long-term trend from the
primary and secondary minima in each target--sector subset, we
converted the residuals to seconds and plotted the two eclipse types
separately in Figure~\ref{fig:oconnell_oc}. Their residuals vary in
opposite directions, particularly for J011323 in Sector 57. Such
anti-correlated timing variations can result from migrating spots that
distort the eclipse profiles \citep{Tran2013,Meng2024}, supporting the
spot-related interpretation of the observed asymmetry.

	\subsection{Absolute Parameter Estimation}
Using PHOEBE, we obtained photometric solutions for all 11 systems. Because
the light curves exhibit total eclipses, the eclipse geometry strongly
constrains the photometric parameters, especially the mass ratio and orbital
inclination \citep{Pribulla2003,TerrellWilson2005,Li2021}. Due to the lack of
radial-velocity data, the absolute physical parameters of the individual
components need to be estimated indirectly.

	For binaries without radial velocities, absolute parameters are commonly
estimated either from Gaia parallax or from an empirical period--semimajor-axis
($P$--$a$) relation. The Gaia-based method requires reliable astrometry
(RUWE $<1.4$) and low extinction ($A_V \lesssim 0.4$; \citealt{Poro2024a}).
Most of our targets fail at least one of these conditions: five systems have
$A_V>0.4$ (Table~\ref{tab:basic_information}), and J085750 has a large RUWE of
7.847. We therefore adopted the empirical $P$--$a$ relation of
\citet{Li2022ELMRCBI} to estimate the orbital semimajor axis:

	\begin{equation}
		a(R_\odot) = 0.501\pm0.063 + 5.621\pm0.138\times P(\mathrm{days}).
	\end{equation}

	The radii, masses, and luminosities of the two components were then calculated from

	\begin{equation}
		\left\{
		\begin{aligned}
			R_i &= a\,r_i, \\
			M_{\mathrm{total}} &= M_1 + M_2 = \frac{0.0134\,a^3}{P^2}, \\
			\frac{L_i}{L_\odot} &= \left(\frac{T_i}{T_\odot}\right)^4\left(\frac{R_i}{R_\odot}\right)^2.
		\end{aligned}
		\right.
	\end{equation}

Here $R_i$, $M_i$, and $L_i$ are the radius, mass, and luminosity of each
component, respectively. The resulting absolute parameters and their
uncertainties are listed in Table~\ref{tab:binary}. Because these parameters are
derived from an empirical $P$--$a$ relation rather than from radial-velocity
measurements, they should be regarded as empirical estimates of the component
masses and radii rather than dynamical measurements. They are also subject to additional systematic uncertainties beyond the formal errors listed in the table.

	\begin{deluxetable*}{lllllllll}[ht!]
		\tabletypesize{\scriptsize}
		\tablewidth{0pt}
		\tablecaption{Absolute parameters of the 11 targets \label{tab:binary}}
		\tablehead{
			\colhead{Target} & \colhead{$a$ ($R_\odot$)} & \colhead{$M_1$ ($M_\odot$)} & \colhead{$M_2$ ($M_\odot$)} & \colhead{$R_1$ ($R_\odot$)} & \colhead{$R_2$ ($R_\odot$)} & \colhead{$L_1$ ($L_\odot$)} & \colhead{$L_2$ ($L_\odot$)} & \colhead{$\frac{\log(R_1/R_2)}{\log(M_1/M_2)}$}
		}
		\startdata
		J002747 & $2.306 \pm 0.077$ & $1.438 \pm 0.145$ & $0.155 \pm 0.020$ & $1.361 \pm 0.048$ & $0.528 \pm 0.024$ & $1.779 \pm 0.134$ & $0.256 \pm 0.026$ & $0.425 \pm 0.023$ \\
		J003357 & $2.326 \pm 0.077$ & $1.469 \pm 0.147$ & $0.131 \pm 0.019$ & $1.409 \pm 0.050$ & $0.512 \pm 0.025$ & $2.316 \pm 0.217$ & $0.326 \pm 0.040$ & $0.419 \pm 0.025$ \\
		J004106 & $2.470 \pm 0.079$ & $1.462 \pm 0.142$ & $0.184 \pm 0.024$ & $1.445 \pm 0.049$ & $0.613 \pm 0.026$ & $2.630 \pm 0.190$ & $0.494 \pm 0.049$ & $0.414 \pm 0.024$ \\
		J011323 & $2.202 \pm 0.076$ & $1.401 \pm 0.145$ & $0.161 \pm 0.023$ & $1.277 \pm 0.046$ & $0.500 \pm 0.022$ & $1.545 \pm 0.115$ & $0.253 \pm 0.025$ & $0.434 \pm 0.026$ \\
		J024047 & $2.427 \pm 0.079$ & $1.466 \pm 0.143$ & $0.166 \pm 0.022$ & $1.413 \pm 0.049$ & $0.551 \pm 0.023$ & $2.733 \pm 0.204$ & $0.404 \pm 0.039$ & $0.432 \pm 0.023$ \\
		J025641 & $2.312 \pm 0.077$ & $1.464 \pm 0.147$ & $0.132 \pm 0.019$ & $1.396 \pm 0.050$ & $0.506 \pm 0.025$ & $3.887 \pm 0.287$ & $0.463 \pm 0.051$ & $0.421 \pm 0.025$ \\
		DK Ari & $2.236 \pm 0.076$ & $1.440 \pm 0.148$ & $0.132 \pm 0.020$ & $1.360 \pm 0.049$ & $0.510 \pm 0.027$ & $3.484 \pm 0.278$ & $0.452 \pm 0.053$ & $0.411 \pm 0.027$ \\
		J040106 & $2.105 \pm 0.074$ & $1.440 \pm 0.153$ & $0.095 \pm 0.017$ & $1.309 \pm 0.050$ & $0.412 \pm 0.024$ & $1.676 \pm 0.137$ & $0.190 \pm 0.024$ & $0.425 \pm 0.030$ \\
		J085750 & $2.886 \pm 0.086$ & $1.602 \pm 0.144$ & $0.187 \pm 0.024$ & $1.700 \pm 0.054$ & $0.693 \pm 0.029$ & $4.334 \pm 0.283$ & $0.763 \pm 0.074$ & $0.418 \pm 0.025$ \\
		J091721 & $2.428 \pm 0.079$ & $1.435 \pm 0.140$ & $0.197 \pm 0.024$ & $1.418 \pm 0.048$ & $0.629 \pm 0.027$ & $2.947 \pm 0.209$ & $0.535 \pm 0.052$ & $0.409 \pm 0.022$ \\
		J232802 & $2.817 \pm 0.085$ & $1.593 \pm 0.145$ & $0.172 \pm 0.021$ & $1.688 \pm 0.054$ & $0.676 \pm 0.030$ & $3.643 \pm 0.242$ & $0.584 \pm 0.060$ & $0.411 \pm 0.023$ \\
		\enddata
	\end{deluxetable*}

	\section{Spectroscopic Investigation}

Chromospheric activity in late-type binaries can be diagnosed from excess
emission in activity-sensitive lines such as H$\alpha$, H$\beta$, H$\gamma$,
Ca\,II H\&K, and Ca\,II IRT \citet{Montes1995}. Because the observed
LAMOST spectra contain both photospheric and chromospheric components, we
applied the spectral-subtraction method to isolate possible excess emission
\citep{Barden1984,Barden1985,Li2025b,Zheng2021}.

\begin{figure*}[ht!]
		\centering
		\includegraphics[width=\textwidth]{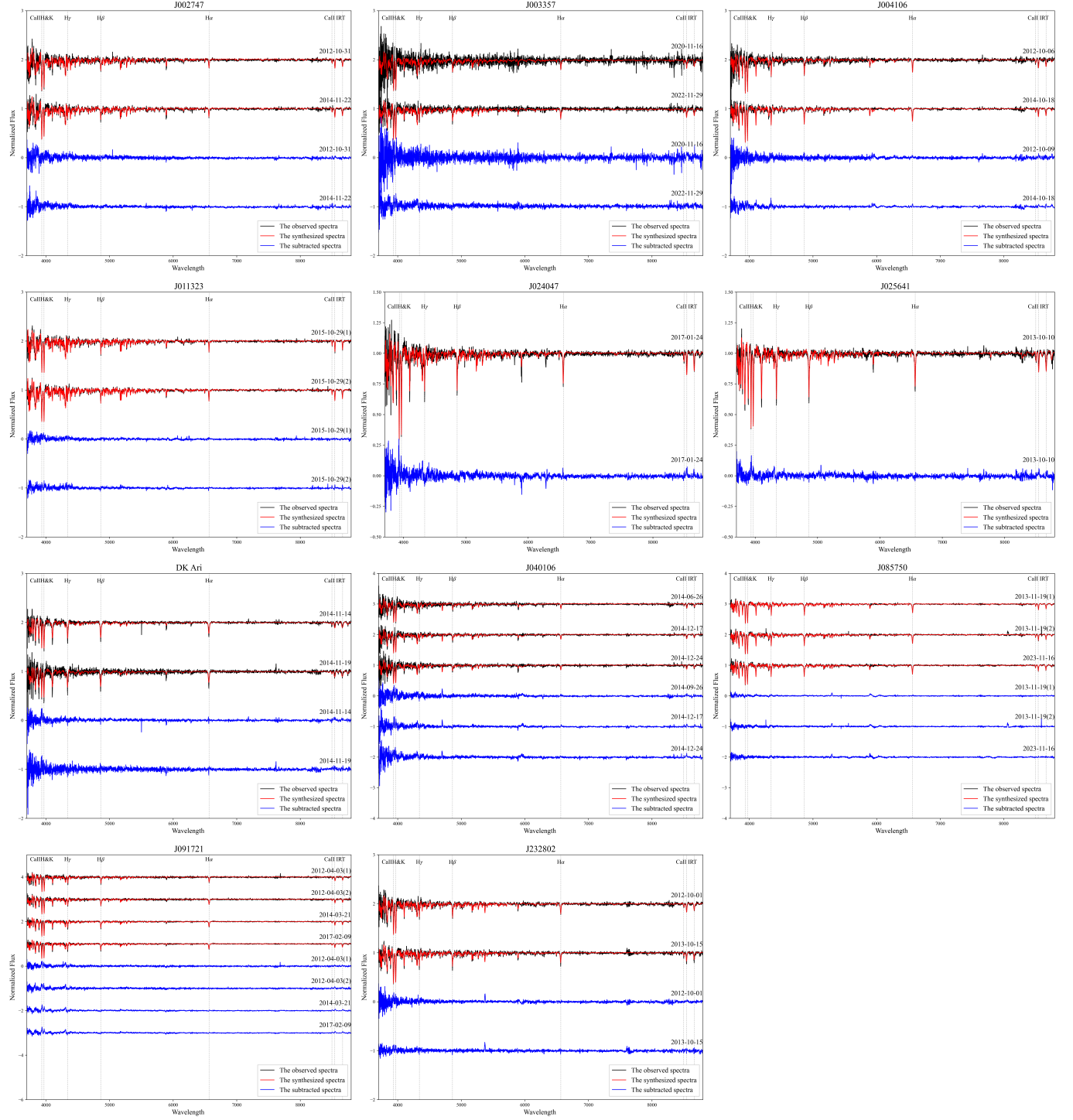}
		\caption{Spectral analysis for the 11 targets. The observed, synthesized, and subtracted spectra are shown for each target, with adjacent spectra shifted vertically by one flux unit. The activity-sensitive lines examined in this work are marked by dashed vertical lines.}
		\label{fig:spc_all}
	\end{figure*}

For the two components of each target, we selected inactive stellar spectra from the catalog of \citet{Zhang2021} as reference templates. In addition to requiring $|\Delta T|<100$ K and $SNR_g>300$, we required $|\Delta\log g|<0.2$ dex and $|\Delta[\mathrm{Fe/H}]|<0.5$ dex. STARMOD \citep{Barden1985} was then used to construct a synthetic inactive composite spectrum. The luminosity ratio was fixed from the PHOEBE solution, whereas the radial velocities and rotational velocities were automatically optimized during the spectral synthesis. The observed, synthesized, and subtracted spectra are shown in Figure~\ref{fig:spc_all}.

	We used the subtracted H$\alpha$ profile as the primary diagnostic of
detectable chromospheric excess emission. H$\alpha$ excess emission is detected
in J024047, J040106, J003357, J004106, J085750, and J091721, but not in
J011323, DK Ari, J025641, J232802, or J002747. Thus, J011323 and J002747 show
prominent O'Connell effects without detectable H$\alpha$ excess in the
available LAMOST spectra. Because the photometry and spectroscopy are not
contemporaneous, this comparison cannot establish whether the O'Connell effect
is directly linked to chromospheric activity.

\section{Orbital Period Variation}
\label{OmC}
To investigate orbital-period variations, we compiled times of minimum light
from our own observations and from the TESS, ZTF, SuperWASP, ASAS-SN, and CRTS
surveys. For well-sampled continuous light curves, such as those from TESS and
SuperWASP, the minima were measured directly with the K--W method
\citep{KweeWoerden1956}. For sparsely sampled survey data, such as those from
ZTF, ASAS-SN, and CRTS, we applied the period-shift method of \citet{Li2020}.
All timings were placed on the BJD$_{\mathrm{TDB}}$ system using the
time-conversion tool of \citet{Eastman2010}. The collected minima and their
data sources are listed in Table~\ref{tab:oc}.

\begin{figure*}[ht!]
		\centering
		\includegraphics[width=\textwidth]{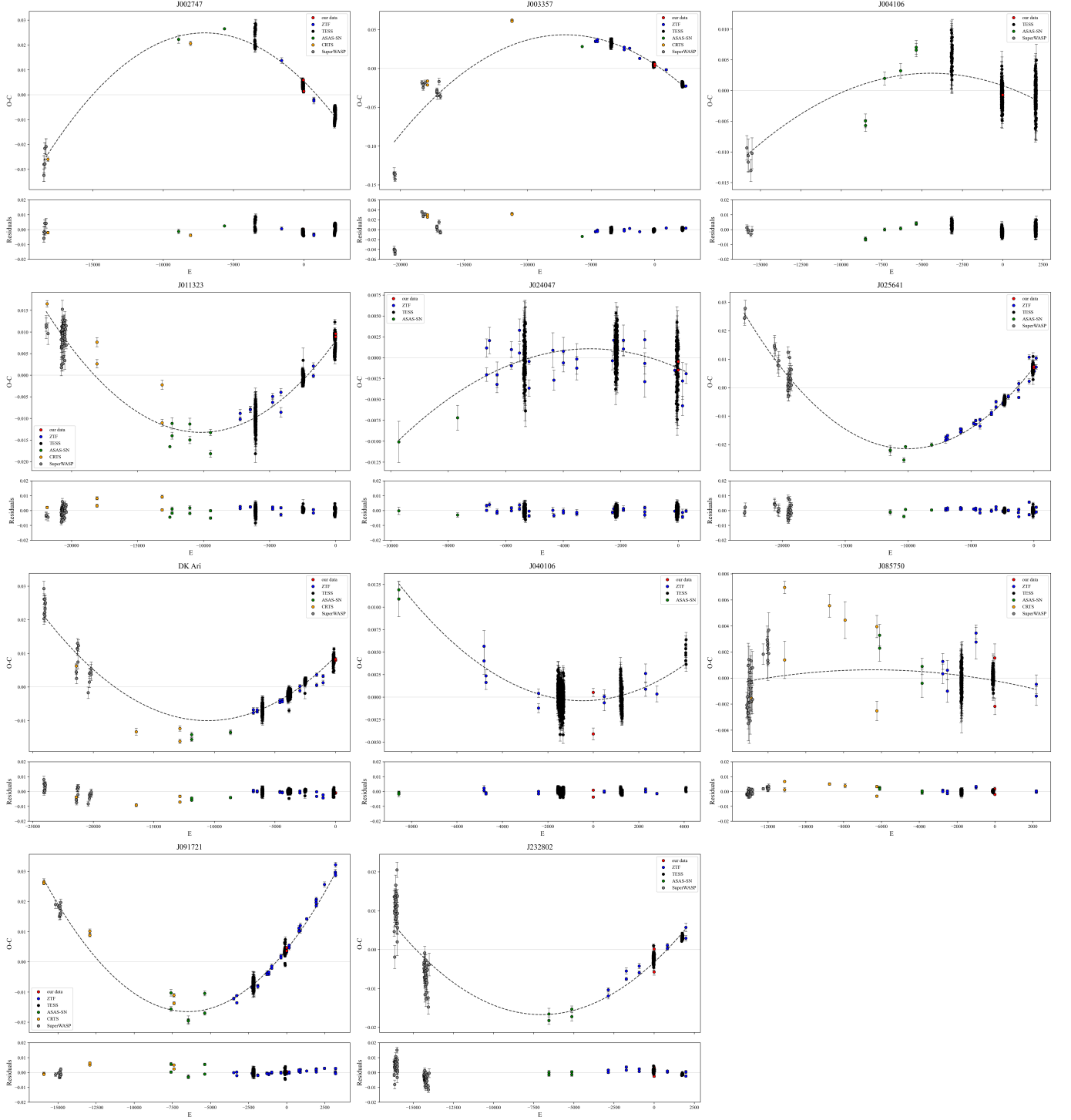}
		\caption{O$-$C diagrams of the 11 targets. The dashed curves represent the fitted ephemerides, while the lower panels show the fitting residuals.}
		\label{fig:omc_all}
	\end{figure*}

For each system, we adopted the primary minimum from our observations
as the initial epoch and the ASAS-SN period as the initial period. The
O$-$C values were first calculated from the linear ephemeris
\begin{equation}
	BJD = BJD_0 + P \times E,
\end{equation}
where $E$ is the cycle number. The corresponding O$-$C diagrams are shown in
Figure~\ref{fig:omc_all}.

After correcting the initial epoch and period with a linear fit, we fitted a
quadratic ephemeris to quantify the secular period variation:
\begin{equation}
	O-C = \Delta T_0 + \Delta P \times E
	+ \frac{\beta}{2}P_0 \times E^2,
\end{equation}
where $\beta$ is the secular period-change rate. The fitted parameters are
summarized in Table~\ref{tab:ephem}.

The quadratic fits indicate secular period increases in J011323,
DK Ari, J025641, J040106, J232802, and J091721, and decreases in
J024047, J003357, J002747, J004106, and J085750. J085750 shows the
weakest decrease, with
$\beta=(-0.35\pm0.11)\times10^{-7}\,\mathrm{d\,yr^{-1}}$. Because the
available baselines and sampling densities remain limited for some
targets, continued eclipse-timing observations are required to test
for higher-order or cyclic variations.

	\begin{deluxetable}{lcccccc}[ht!]
		\tabletypesize{\scriptsize}
		\tablewidth{0pt}
		\tablecaption{Times of minimum and O$-$C values for the 11 targets \label{tab:oc}}
		\tablehead{
			\colhead{Target} & \colhead{Min. (BJD$_{\rm TDB}$)} & \colhead{Error} & \colhead{$E$} & \colhead{O-C} & \colhead{Residuals} &\colhead{Data source}
		}
		\startdata
		J002747 & 2459885.35097 & 0.00038 & 0 & 0.00140 & -0.00398 & NEXT \\
		& 2459874.27545 & 0.00049 & -34.5 & 0.00593 & 0.00036 & NEXT \\
		J003357 & 2459889.87763 & 0.00049 & 0 & 0.00342 & -0.00196 & SPM84 \\
		& 2459889.71759 & 0.00072 & -0.5 & 0.00571 & 0.00032 & SPM84 \\
		J004106 & 2459883.25299 & 0.00040 & 0 & -0.00073 & -0.00151 & NEXT \\
		& 2458764.86127 & 0.00224 & -3192 & 0.00318 & 0.00053 & TESS \\
		J011323 & 2460613.22198 & 0.00021 & 0 & 0.00960 & 0.00172 & XL85 \\
		& 2460613.06996 & 0.00034 & -0.5 & 0.00890 & 0.00101 & XL85 \\
		\enddata
		\tablecomments{This table is available in its entirety in machine-readable form in the online version of this paper.}
	\end{deluxetable}

\begin{deluxetable*}{lrcrcrrcc}[ht!]
	\tabletypesize{\scriptsize}
	\tablewidth{0pt}
	\tablecaption{Ephemeris parameters, conservative mass-transfer rates, and orbital angular momenta of the targets \label{tab:ephem}}
	\tablehead{
		\colhead{Target} & \colhead{$\Delta P$ ($10^{-6}$\,d)} & \colhead{Corrected $P$} & \colhead{$\Delta Min.$ ($10^{-3}$\,d)} & \colhead{Corrected $Min.$} & \colhead{$\beta$ ($10^{-7}$\,d\,yr$^{-1}$)} & \colhead{$\dot{M}_1$ ($10^{-8}\,M_\odot$\,yr$^{-1}$)} & \colhead{$J_{\rm orb}$ ($10^{51}$\,cgs)} & \colhead{$J_{\rm spin}/J_{\rm orb}$}
	}
	\startdata
	J002747 & $1.64 \pm 0.16$ & 0.3211606(2) & $-1.40 \pm 0.68$ & 2459885.34956(7) & $-8.91 \pm 0.13$ & $-16.10 \pm 2.23$ & $1.62 \pm 0.30$ & $0.15 \pm 0.03$ \\
	J003357 & $-1.40 \pm 0.29$ & 0.3246465(3) & $-3.42 \pm 1.29$ & 2459889.87420(7) & $-17.20 \pm 0.24$ & $-25.34 \pm 3.86$ & $1.40 \pm 0.27$ & $0.18 \pm 0.04$ \\
	J004106 & $-3.93 \pm 0.06$ & 0.3503746(6) & $0.73 \pm 0.18$ & 2459883.25371(3) & $-2.14 \pm 0.16$ & $-4.29 \pm 0.67$ & $1.99 \pm 0.35$ & $0.13 \pm 0.03$ \\
	J011323 & $1.49 \pm 0.06$ & 0.3026268(6) & $-9.60 \pm 0.42$ & 2460613.21238(0) & $4.91 \pm 0.06$ & $9.85 \pm 1.53$ & $1.62 \pm 0.31$ & $0.15 \pm 0.03$ \\
	J024047 & $-0.63 \pm 0.06$ & 0.3426488(6) & $1.48 \pm 0.20$ & 2460635.28121(8) & $-5.28 \pm 0.55$ & $-9.59 \pm 1.67$ & $1.79 \pm 0.32$ & $0.14 \pm 0.03$ \\
	J025641 & $-1.18 \pm 0.07$ & 0.3221726(7) & $-6.98 \pm 0.44$ & 2460648.26739(3) & $6.43 \pm 0.06$ & $9.63 \pm 1.46$ & $1.40 \pm 0.27$ & $0.18 \pm 0.04$ \\
	DK Ari  & $0.35 \pm 0.05$ & 0.3087267(5) & $-7.94 \pm 0.26$ & 2460668.17530(8) & $4.04 \pm 0.03$ & $6.36 \pm 1.01$ & $1.37 \pm 0.27$ & $0.19 \pm 0.04$ \\
	J040106 & $0.51 \pm 0.04$ & 0.2852770(4) & $-0.52 \pm 0.07$ & 2459894.76510(0) & $5.09 \pm 0.27$ & $6.05 \pm 1.17$ & $0.97 \pm 0.21$ & $0.26 \pm 0.07$ \\
	J085750 & $-0.26 \pm 0.02$ & 0.4242530(4) & $2.17 \pm 0.08$ & 2459634.17907(5) & $-0.35 \pm 0.11$ & $-0.58 \pm 0.20$ & $2.30 \pm 0.39$ & $0.13 \pm 0.03$ \\
	J091721 & $0.53 \pm 0.15$ & 0.3428101(2) & $-3.65 \pm 0.53$ & 2459631.13975(0) & $10.50 \pm 0.09$ & $23.26 \pm 2.99$ & $2.08 \pm 0.36$ & $0.13 \pm 0.03$ \\
	J232802 & $3.65 \pm 0.05$ & 0.4120771(5) & $-0.14 \pm 0.31$ & 2459850.37908(7) & $4.85 \pm 0.17$ & $7.56 \pm 1.03$ & $2.09 \pm 0.35$ & $0.14 \pm 0.03$ \\
	\enddata
	\tablecomments{Corrected Min. are given in BJD$_{\rm TDB}$. The values of $\dot{M}_1$ are diagnostic estimates calculated under the conservative mass-transfer assumption.}
\end{deluxetable*}
	
Secular period increases may result from mass transfer from the less
massive component to the more massive component, whereas decreases may
be produced by reverse mass transfer, angular-momentum loss (AML), or
both. Under the conservative mass-transfer assumption, the corresponding
diagnostic rate can be estimated as \citep{Kwee1958}:

	\begin{equation}
		\frac{dM_1}{dt} = \frac{M_1 M_2}{3P(M_1-M_2)}\times\frac{dP}{dt},
		\label{eq:mdot}
	\end{equation}
	
where $dP/dt$ corresponds to $\beta$ in Table~\ref{tab:ephem}. The resulting rates are listed in Table~\ref{tab:ephem}; they are
diagnostic estimates under the conservative assumption, rather than the
long-term mass-transfer rates adopted in the evolutionary models below.

For the five systems with negative $\beta$, we estimated the possible
AML contribution using the magnetic-braking prescription of
\citet{GuinanBradstreet1988}:

	\begin{equation}
		\dot{P}_{\rm AML} \approx -1.1\times10^{-8}\cdot\frac{(1+q)^2}{q}\cdot\frac{k_1^2 M_1 R_1^4+k_2^2 M_2 R_2^4}{(M_1+M_2)^{5/3}P^{7/3}}.
		\label{eq:paml}
	\end{equation}
The gyration radius of star 1 was estimated from the relations for rotating
and tidally distorted ZAMS stars given by \citet{arbutina2024critical}, yielding
a mean value of $k_1^2\simeq0.04$ for our sample. Star 2 was treated as a fully
convective star with $k_2^2=0.205$ \citep{Arbutina2007}. The resulting $\dot{P}_{\rm AML}$ values are
$-1.48$, $-2.08$, $-1.61$, $-1.40$, and
$-1.55\times10^{-7}\,\mathrm{d\,yr^{-1}}$ for J024047, J003357, J002747,
J004106, and J085750, respectively. For the first four systems, these
values correspond to approximately 28\%, 12\%, 18\%, and 66\% of the
observed period decreases.
For J085750, the nominal AML rate exceeds its weak observed period
decrease, so a simple fractional comparison is not meaningful. The
difference may reflect limitations in the data quality, or unmodeled
mass transfer from star 2 to star 1 that partly offsets the AML-driven
period decrease. Overall, the decreasing periods probably result from a
combination of AML and mass transfer. Because the classical prescription neglects magnetic-braking saturation and overestimates the AML rate by an order of magnitude or more in the contact-binary regime \citep{ElBadry2022,Fabry2025b}, the derived rates should be regarded as overestimates.
Gravitational radiation also produces unavoidable AML, but its
contribution at the periods and masses of this sample is expected to be
smaller than the classical magnetic-braking estimate and is not
included in the numerical comparison above\citep{Li2004}.

\section{Discussion and Conclusions}
\label{dis}

The photometric solutions confirm all 11 targets as ELMRCBs
($q<0.15$). Following \citet{Li2022ELMRCBI}, J011323 and J024047 are
medium-contact systems ($0.25\leq f<0.50$), while the remaining nine are
deep-contact systems ($f\geq0.50$); none is in shallow contact. Based on the
nominal corrected component temperatures, J003357, J004106, J011323, J040106,
and J085750 are W-subtype systems, while the others are A-subtype systems. The TESS light
curves of J011323 and J002747 show time-dependent O'Connell effects, with the
asymmetry reversing within a single sector. Spectral subtraction reveals excess
chromospheric emission in J024047, J040106, J003357, J004106, J085750, and
	J091721. The O$-$C analysis indicates secular period increases in six systems
	and secular period decreases in five systems, with J085750 showing the weakest
	decreasing trend.

\begin{figure*}[ht!]
		\centering
		\includegraphics[width=0.95\textwidth]{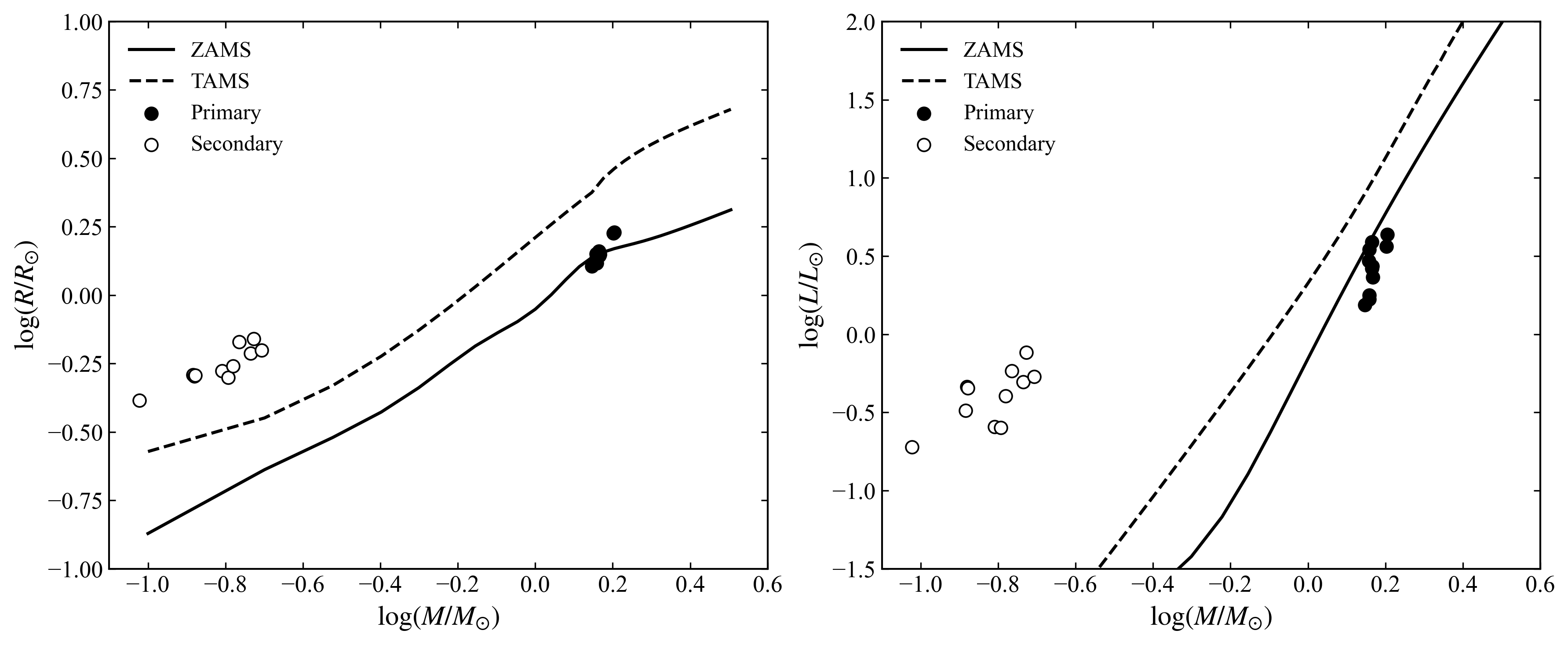}
		\caption{Logarithmic M-L and M-R diagrams of the target components.}
		\label{fig:zams_tams}
	\end{figure*}

To examine the evolutionary states of the components, we generated
zero-age and terminal-age main-sequence (ZAMS and TAMS) reference
relations with the binary-star evolution code of
\citet{hurley2002BSE}. As shown in Figure~\ref{fig:zams_tams}, the
more massive components lie close to the ZAMS, whereas the less
massive components lie above the TAMS and are oversized and
overluminous for their present masses. For two ZAMS components,
$\log(R_1/R_2)/\log(M_1/M_2)$ is expected to be approximately 0.64
\citep{Allen1973}, compared with 0.41--0.43 for our targets (Table~\ref{tab:binary}), further suggesting that the secondary components are
larger than normal main-sequence stars of the same mass. 
These properties are consistent with energy
transfer in contact binaries
\citep{Flannery1976,Li2004,YakutEggleton2005,Fabry2025b}
and should not be interpreted as direct nuclear-age indicators for
the less massive components.

We calculated the current orbital angular momentum \(J_{\rm orb}\) for all targets using the standard expression for circular orbits:
\[
J_{\rm orb} = 1.24 \times 10^{52}\,
M_{\rm tot}^{5/3}\, P^{1/3}\,
\frac{q}{(1+q)^2}\]
The results are listed in Table~\ref{tab:ephem}.

We then evaluated the present dynamical stability of the systems using the Darwin criterion following \citet{YangQian2015}:
\[
\frac{J_{\mathrm{spin}}}{J_{\mathrm{orb}}}= \frac{1+q}{q}\left[(k_1r_1)^2+(k_2r_2)^2q\right]
\]
where the gyration radii \(k_1\) and \(k_2\) are adopted consistently with the angular-momentum-loss calculation. All targets have \(J_{\mathrm{spin}}/J_{\mathrm{orb}}<1/3\) and are therefore dynamically stable within the quoted uncertainties. J040106 has the largest value of \(0.26\pm0.07\), making it the closest system to the Darwin-instability limit.

{To examine whether short-period detached binaries can evolve to the observed extremely low mass ratios, short periods, and low orbital angular momenta, we constructed a semi-analytical model for contact binary evolution from a dynamical perspective.} The model assumes synchronous rotation and circular orbits, enforcing \(\omega_{\rm spin} = \omega_{\rm orb}\) at each time step. AML is driven by saturated magnetic braking (MB; \citealp{Sills2000}) and gravitational radiation (GR; \citealp{Landau1975}) from the quadrupole component. The activation of MB is determined by stellar mass: for fully convective stars with \(M \lesssim 0.30\,M_\odot\), the MB strength is set to zero; in the mass range \(1.2\)--\(1.35\,M_\odot\), the MB efficiency is smoothly reduced from 1.0 to 0.2. Both conservative and non-conservative mass transfer are tested, and wind mass loss follows \citet{Reimers1975}.

The Roche lobes of the two components are determined following \citet{Eggleton1983}. Radii and moments of inertia are interpolated from the non-rotating PARSEC v2.0 single-star grids \citep{Costa2025PARSEC} as functions of mass and nuclear reaction progress, with metallicities of \(Z=0.008\) and \(0.017\). Since these grids cannot reproduce the current thermal expansion state of low-mass secondary stars, and we do not include energy transfer, we adopt an empirical scaling for the secondary radius after contact. Following the statistical results of \citet{Li2024}, we set \(R_s = R_p \, q^{0.4}\) as a reference for the secondary radius.

The evolutionary process is divided into three phases following \citet{Stepien2006} and \citet{StepienKiraga2015}. In Phase~I, the two components evolve independently from the ZAMS, the more massive star gradually expands due to nuclear evolution until it fills its Roche lobe. In Phase~II, the more massive star transfers material to the accretor in a rapid mass-exchange phase, with the mass-transfer rate set to \(7\times10^{-9}\,M_\odot\,\mathrm{yr}^{-1}\), until the less massive star also fills its Roche lobe. Phase~III is a long-lived contact phase, during which the mass-transfer rate is dynamically adjusted according to the relative overflow and the radius--mass-ratio relation, remaining at approximately \(3\times10^{-10}\,M_\odot\,\mathrm{yr}^{-1}\).

\begin{figure*}[ht!]
		\centering
		\includegraphics[width=0.95\textwidth]{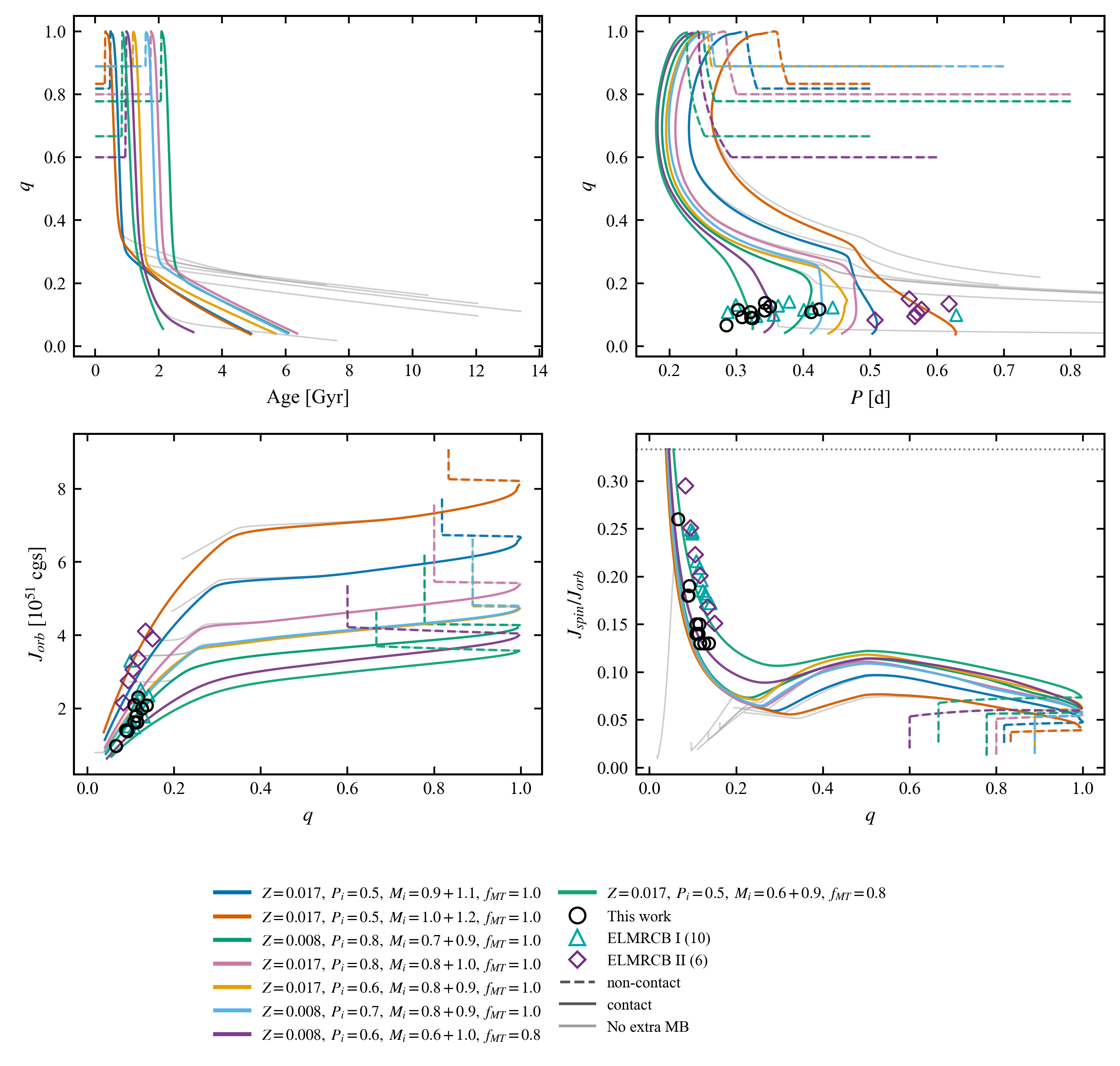}
		\caption{Representative evolutionary tracks from the modified detached binary
		formation channel for the ELMRCB sample.}
		\label{fig:ccbm_tracks}
	\end{figure*}

A wide range of initial parameter combinations was explored through orbital integration. Our tests show that when AML is driven solely by saturated MB and GR, with the mass-transfer rates adopted above, the AML efficiency is insufficient to reduce \(J_{\rm orb}\) to the observed level of ELMRCBs before the accretor reaches the TAMS. If energy transfer is further introduced to enhance the mass-transfer rate, the AML efficiency becomes even more weakened in the late stages, leading to rapid orbital expansion. In the colored tracks shown in Figure~\ref{fig:ccbm_tracks}, we impose a residual MB strength of 20\% for stars with \(M > 1.35\,M_\odot\) to regulate the late-stage evolution, and we find that an appropriate level of non-conservative mass transfer is also required to cover the short-period end. For comparison, tracks with the same initial parameters but without this additional MB are shown in gray. Even with both the residual MB strength and non-conservative mass transfer included, the colored tracks still fail to reproduce the shortest-period targets such as J040106. This suggests that future complete modeling of ELMRCB evolution must incorporate an AML mechanism that is structurally self-consistent and capable of releasing sufficient angular momentum during the contact phase, {unless such a requirement is alleviated by a lower mass-transfer rate or by including energy transfer, which would alter the radius evolution.}

 It should be noted that the retained 20\% MB strength adopted above does not represent a real magnetic braking remnant, but rather a parameterized proxy for additional AML mechanisms not covered by the standard saturated MB and GR prescriptions. The MB strength, the AML by mass outflow, and other factors collectively influence the contraction or expansion of the orbital radius. Therefore, the results presented here should be regarded as one possible combination of parameters, rather than the unique pathway to forming ELMRCBs. All colored tracks terminate via the Darwin instability, with terminal mass ratios of \(q=0.037\)--\(0.055\). Because our model does not include \(L_2\) overflow, we cannot rule out that actual systems would instead lose mass through \(L_2\) before reaching the Darwin limit. Given the semi-analytical nature of the model and the absence of complete simulations of energy transfer and radius evolution, the above discussion is limited to a dynamical feasibility test of evolutionary pathways, and should not be interpreted as a quantitative reconstruction of the evolutionary history of individual targets.

	\begin{acknowledgments}
		
	We thank the anonymous reviewer for insightful comments and constructive suggestions, which have significantly improved the quality of this manuscript. 
	This work is supported by the National Natural Science Foundation of China (NSFC; No. 12273018), the Taishan Scholars Young Expert Program of Shandong Province, the Qilu Young Researcher Project of Shandong University, the Young Data Scientist Project of the National Astronomical Data Center, the Cultivation Project for LAMOST Scientific Payoff and Research Achievement of CAMS-CAS, the International Centre of Supernovae (ICESUN), Yunnan Key Laboratory of Supernova Research (No. 202505AV340004), and the Key Undergraduate Teaching Reform Research Project of Shandong Province (Grant No. Z2023200). The calculations in this work were carried out at the Supercomputing Center of Shandong University, Weihai.

	This work made use of data from LAMOST (Large Sky Area Multi-Object Fiber Spectroscopic Telescope, also known as the Guoshoujing Telescope; \url{https://cstr.cn/31118.02.LAMOST}), a Chinese national mega-science facility operated by National Astronomical Observatories, Chinese Academy of Sciences.

	We acknowledge the support of the staff of the Xinglong 85cm telescope. This work was partially supported by National Astronomical Observatories, Chinese Academy of Sciences.

	This paper includes data collected by TESS. Funding for TESS is provided by NASA's Science Mission Directorate. TESS data in this paper were obtained from the Mikulski Archive for Space Telescopes (MAST) at the Space Telescope Science Institute.

	This work made use of data from the European Space Agency (ESA) mission Gaia \citep{10.5270/esa-1ugzkg7}, processed by the Gaia Data Processing and Analysis Consortium (DPAC; \url{https://www.cosmos.esa.int/web/gaia/dpac/consortium}).

	This paper makes use of data from ASAS-SN. ASAS-SN is funded in part by the Gordon and Betty Moore Foundation through grant numbers GBMF5490 and GBMF10501 to the Ohio State University, and also funded in part by the Alfred P. Sloan Foundation grant No G-2021-14192.
	
	This paper makes use of data products from the Two Micron All Sky Survey (2MASS; \citealt{10.26131/irsa2}), a joint project of the University of Massachusetts and the Infrared Processing and Analysis Center/California Institute of Technology, funded by NASA and the National Science Foundation.

	This paper makes use of ZTF data from IRSA \citep{10.26131/irsa598} and CRTS data \citep{10.26093/cds/vizier.16960870}. This paper also makes use of data from the SuperWASP public archive \citep{Butters2010,10.26133/nea9}, as provided by the WASP consortium, and computational resources supplied by the project e-Infrastruktura CZ (e-INFRA CZ LM2018140), supported by the Ministry of Education, Youth and Sports of the Czech Republic.
\end{acknowledgments}

\appendix
\restartappendixnumbering
	\section{TESS Light-curve Analysis}\label{app:tess}

\begin{figure*}[ht!]
		\centering
		\includegraphics[width=\textwidth]{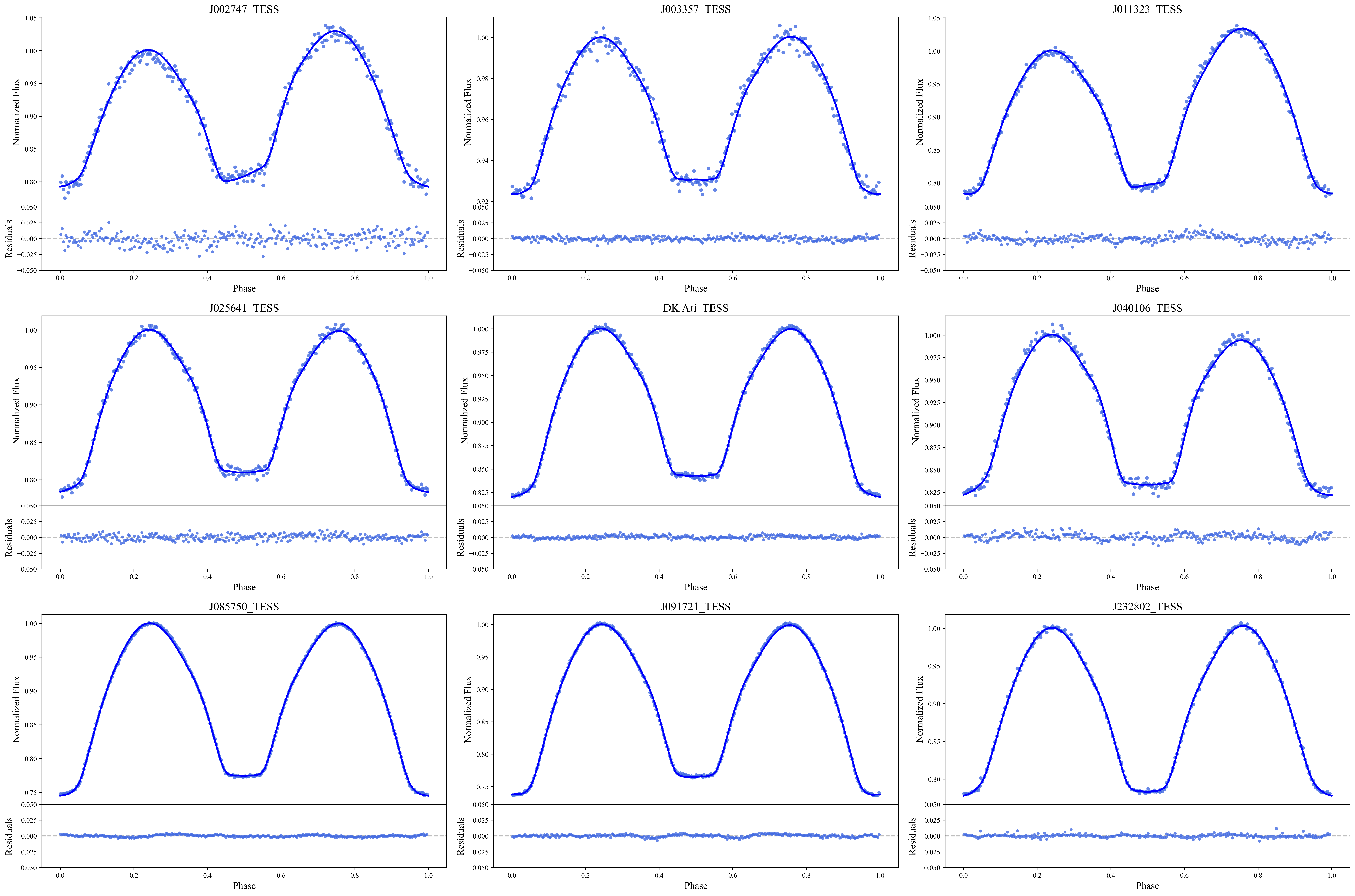}
		\caption{Comparison between the observed and synthetic light curves for the nine targets.}
		\label{fig:tess_lc_appendix}
	\end{figure*}

	Figure~\ref{fig:tess_lc_appendix} shows the TESS light-curve solutions with CBLA for the nine targets with usable TESS data, and Table~\ref{tab:tess_results} lists the corresponding parameters. The fixed temperatures of star 1 are those in Table~\ref{tab:results}. After removing low-SNR measurements, we phase-folded the 2 or 10 minute cadence data and binned each light curve into 200 points. Because one TESS pixel subtends about $21\arcsec$ (\citealt{Brasseur2019Astrocut}), non-zero third-light terms were included for J011323 and J003357 based on the amplitude differences between the TESS and ground-based light curves. Using the TESS-cont tool \citep{Cast2024}, we estimated aperture contamination fractions of 35.57--65.75\% for J003357 and 17.12--40.21\% for J011323. The fitted $l_3$ values fall within the corresponding contamination ranges. J024047 and J004106 were not included because reliable TESS data could not be extracted. For the remaining targets, the TESS solutions are generally consistent with the ground-based solutions, supporting the reliability of the adopted photometric parameters.

	\begin{table*}[ht!]
		\centering
		\caption{TESS light-curve solutions for the nine targets \label{tab:tess_results}}
		\setlength{\tabcolsep}{4pt}
		\scriptsize
		\begin{tabular}{lcccccc}
			\hline\hline
			Target & $q$ & $i$ (deg) & $T_2$ (K) & $f$ & $L_1/L_T$ & $L_2/L_T$ \\
			\hline
			J002747 & $0.0985_{-0.0002}^{+0.0003}$ & $78.34_{-0.08}^{+0.08}$ & $5887_{-4}^{+4}$ & $0.565_{-0.007}^{+0.007}$ & $0.8772_{-0.0002}^{+0.0002}$ & $0.1228_{-0.0002}^{+0.0002}$ \\
			J003357 & $0.0924_{-0.0001}^{+0.0002}$ & $76.37_{-0.06}^{+0.07}$ & $5820_{-5}^{+7}$ & $0.320_{-0.003}^{+0.002}$ & $0.3618_{-0.0001}^{+0.0001}$ & $0.0404_{-0.0001}^{+0.0001}$ \\
			J011323 & $0.1014_{-0.0003}^{+0.0004}$ & $74.80_{-0.08}^{+0.06}$ & $6183_{-3}^{+2}$ & $0.554_{-0.004}^{+0.003}$ & $0.6150_{-0.0010}^{+0.0010}$ & $0.1133_{-0.0003}^{+0.0003}$ \\
			J025641 & $0.0938_{-0.0001}^{+0.0001}$ & $84.73_{-0.09}^{+0.06}$ & $6597_{-2}^{+2}$ & $0.524_{-0.002}^{+0.002}$ & $0.8910_{-0.0001}^{+0.0001}$ & $0.1090_{-0.0001}^{+0.0001}$ \\
			DK Ari & $0.0853_{-0.0001}^{+0.0001}$ & $78.24_{-0.07}^{+0.07}$ & $6624_{-4}^{+3}$ & $0.611_{-0.004}^{+0.004}$ & $0.9114_{-0.0001}^{+0.0001}$ & $0.0886_{-0.0001}^{+0.0001}$ \\
			J040106 & $0.0673_{-0.0005}^{+0.0004}$ & $79.34_{-0.23}^{+0.19}$ & $5994_{-4}^{+5}$ & $0.295_{-0.015}^{+0.019}$ & $0.9009_{-0.0001}^{+0.0001}$ & $0.0991_{-0.0001}^{+0.0001}$ \\
			J085750 & $0.1185_{-0.0003}^{+0.0003}$ & $78.99_{-0.12}^{+0.11}$ & $6411_{-1}^{+1}$ & $0.638_{-0.006}^{+0.007}$ & $0.8667_{-0.0001}^{+0.0001}$ & $0.1333_{-0.0001}^{+0.0001}$ \\
			J091721 & $0.1229_{-0.0002}^{+0.0002}$ & $81.59_{-0.07}^{+0.06}$ & $6155_{-2}^{+2}$ & $0.674_{-0.004}^{+0.004}$ & $0.8583_{-0.0001}^{+0.0001}$ & $0.1417_{-0.0001}^{+0.0001}$ \\
			J232802 & $0.0960_{-0.0001}^{+0.0001}$ & $74.70_{-0.05}^{+0.05}$ & $6254_{-2}^{+2}$ & $0.704_{-0.003}^{+0.002}$ & $0.8689_{-0.0001}^{+0.0001}$ & $0.1311_{-0.0001}^{+0.0001}$ \\
			\hline
		\end{tabular}

		\vspace{1ex}

		\begin{tabular}{lcrrccc}
			\hline\hline
			Target & $L_3/L_T$ & $\lambda$ (deg) & $r_s$ (deg) & $T_s$ & $r_1$ & $r_2$ \\
			\hline
			J002747 & \nodata & $127.3_{-0.4}^{+0.4}$ & $11.3_{-0.1}^{+0.1}$ & $1.176_{-0.003}^{+0.003}$ & $0.6048_{-0.0004}^{+0.0004}$ & $0.2171_{-0.0001}^{+0.0001}$ \\
			J003357 & $0.5980_{-0.0021}^{+0.0023}$ & $255.8_{-1.5}^{+1.7}$ & $2.7_{-0.2}^{+0.2}$ & $0.724_{-0.009}^{+0.017}$ & $0.5970_{-0.0001}^{+0.0001}$ & $0.2099_{-0.0001}^{+0.0001}$ \\
			J011323 & $0.2720_{-0.0012}^{+0.0014}$ & $184.1_{-0.1}^{+0.1}$ & $9.3_{-0.1}^{+0.1}$ & $1.258_{-0.001}^{+0.001}$ & $0.5951_{-0.0003}^{+0.0002}$ & $0.2264_{-0.0002}^{+0.0003}$ \\
			J025641 & \nodata & $42.8_{-1.0}^{+1.0}$ & $7.4_{-0.1}^{+0.2}$ & $0.927_{-0.004}^{+0.004}$ & $0.6020_{-0.0001}^{+0.0001}$ & $0.2180_{-0.0001}^{+0.0001}$ \\
			DK Ari & \nodata & $170.7_{-1.1}^{+0.9}$ & $8.7_{-0.2}^{+0.2}$ & $0.999_{-0.001}^{+0.001}$ & $0.6120_{-0.0002}^{+0.0002}$ & $0.2010_{-0.0001}^{+0.0001}$ \\
			J040106 & \nodata & $30.4_{-0.2}^{+0.1}$ & $11.5_{-0.2}^{+0.1}$ & $0.904_{-0.005}^{+0.002}$ & $0.6160_{-0.0007}^{+0.0009}$ & $0.1910_{-0.0003}^{+0.0002}$ \\
			J085750 & \nodata & $0.9_{-0.1}^{+0.2}$ & $7.0_{-0.1}^{+0.1}$ & $0.624_{-0.009}^{+0.006}$ & $0.5870_{-0.0003}^{+0.0004}$ & $0.2410_{-0.0001}^{+0.0002}$ \\
			J091721 & \nodata & $0.4_{-0.2}^{+0.3}$ & $11.2_{-0.1}^{+0.1}$ & $0.820_{-0.004}^{+0.005}$ & $0.5860_{-0.0002}^{+0.0002}$ & $0.2457_{-0.0001}^{+0.0001}$ \\
			J232802 & \nodata & $230.5_{-0.7}^{+0.9}$ & $2.3_{-0.1}^{+0.1}$ & $0.708_{-0.005}^{+0.004}$ & $0.6028_{-0.0001}^{+0.0001}$ & $0.2280_{-0.0001}^{+0.0001}$ \\
			\hline
		\end{tabular}
	\end{table*}

\clearpage
\begin{thebibliography}{}
\expandafter\ifx\csname natexlab\endcsname\relax\def\natexlab#1{#1}\fi
\providecommand{\url}[1]{\href{#1}{#1}}
\providecommand{\dodoi}[1]{doi:~\href{http://doi.org/#1}{\nolinkurl{#1}}}
\providecommand{\doeprint}[1]{\href{http://ascl.net/#1}{\nolinkurl{http://ascl.net/#1}}}
\providecommand{\doarXiv}[1]{\href{https://arxiv.org/abs/#1}{\nolinkurl{https://arxiv.org/abs/#1}}}
	
			\bibitem[Allen(1973)]{Allen1973}
Allen, C.~W.\ 1973, Astrophysical Quantities, 3rd edn.\ (London: Athlone Press, University of London)

	\bibitem[Arbutina(2007)]{Arbutina2007}
		Arbutina, B. 2007, MNRAS, 377, 1635, \doi{10.1111/j.1365-2966.2007.11723.x}

	\bibitem[Arbutina(2009)]{Arbutina2009}
		Arbutina, B.\ 2009, MNRAS, 394, 501, \doi{10.1111/j.1365-2966.2008.14332.x}

	\bibitem[Arbutina \& Wadhwa(2024)]{arbutina2024critical}
		Arbutina, B., \& Wadhwa, S. 2024, Serbian Astronomical Journal, 208, 1, \doi{10.2298/SAJ2408001A}

	\bibitem[Balaji et~al.(2015)]{Balaji2015}
		Balaji, B., Croll, B., Levine, A. M., \& Rappaport, S. 2015, MNRAS, 448, 429, \doi{10.1093/mnras/stv031}

	\bibitem[Barden(1984)]{Barden1984}
		Barden, S. C. 1984, BAAS, 16, 893

	\bibitem[Barden(1985)]{Barden1985}
		Barden, S. C. 1985, ApJ, 295, 162, \doi{10.1086/163361}

	\bibitem[Bellm et~al.(2019)]{Bellm2019}
		Bellm, E. C., Kulkarni, S. R., Graham, M. J., et~al. 2019, PASP, 131, 018002, \doi{10.1088/1538-3873/aaecbe}

	\bibitem[Binnendijk(1970)]{Binnendijk1970}
		Binnendijk, L. 1970, Vistas in Astronomy, 12, 217, \doi{10.1016/0083-6656(70)90041-3}

	\bibitem[Brasseur et~al.(2019)]{Brasseur2019Astrocut}
		Brasseur, C. E., Phillip, C., Fleming, S. W., Mullally, S. E., \& White, R. L. 2019, Astrocut: Tools for creating cutouts of TESS images, Astrophysics Source Code Library, ascl:1905.007

	\bibitem[Butters et~al.(2010)]{Butters2010}
		Butters, O. W., West, R. G., Anderson, D. R., et~al. 2010, A\&A, 520, L10, \doi{10.1051/0004-6361/201015655}

\bibitem[Castro-Gonz\'{a}lez et~al.(2024)]{Cast2024}
Castro-Gonz\'{a}lez, A., Lillo-Box, J., Armstrong, D.~J., Acu\~{n}a, L., et~al.\ 2024, \aap, 691, A233, \doi{10.1051/0004-6361/202451656}

	\bibitem[Caton et~al.(2019)]{Caton2019}
Caton, D., Gentry, D.~R., Samec, R.~G., Chamberlain, H., Robb, R., Faulkner, D.~R., \& Hill, R.\ 2019, PASP, 131, 054201, \doi{10.1088/1538-3873/aafb8f}

	\bibitem[Christopoulou \& Papageorgiou(2013)]{ChristopoulouPapageorgiou2013}
		Christopoulou, P.-E., \& Papageorgiou, A. 2013, AJ, 146, 157, \doi{10.1088/0004-6256/146/6/157}

	\bibitem[Conroy et~al.(2020)]{Conroy2020}
		Conroy, K. E., Kochoska, A., Abdelaziz, M., et~al. 2020, ApJS, 250, 34, \doi{10.3847/1538-4365/abb4e2}

	\bibitem[Costa et~al.(2025)]{Costa2025PARSEC}
Costa, G., Shepherd, K. G., Nguyen, C. T., et~al. 2025, A\&A, 694, A193, \doi{10.1051/0004-6361/202452573}

	\bibitem[Cui et~al.(2012)]{Cui2012}
		Cui, X.-Q., Zhao, Y.-H., Chu, Y.-Q., et~al. 2012, RAA, 12, 1197, \doi{10.1088/1674-4527/12/9/003}

	\bibitem[Drake et~al.(2009)]{Drake2009}
		Drake, A. J., Djorgovski, S. G., Mahabal, A., et~al. 2009, ApJ, 696, 870, \doi{10.1088/0004-637X/696/1/870}
		
\bibitem[{{Drake} {et~al.}(2011){Drake}, {Djorgovski}, {Mahabal}, {Beshore},
  {Larson}, {Graham}, {Williams}, {Christensen}, {Catelan}, {Boattini},
  {Gibbs}, {Hill}, \& {Kowalski}}]{10.26093/cds/vizier.16960870}
{Drake}, A.~J., {Djorgovski}, S.~G., {Mahabal}, A., {et~al.} 2011, Catalina
  Real-time Transient Survey (CRTS),  Centre de Donnees Strasbourg (CDS),
  \dodoi{10.26093/cds/vizier.16960870}

	\bibitem[Eastman et~al.(2010)]{Eastman2010}
		Eastman, J., Siverd, R., \& Gaudi, B. S. 2010, PASP, 122, 935, \doi{10.1086/655938}

	\bibitem[El-Badry et~al.(2022)]{ElBadry2022}
		El-Badry, K., Conroy, C., Fuller, J., et~al.\ 2022, MNRAS, 517, 4916, \doi{10.1093/mnras/stac2945}
		
	\bibitem[Eggleton(1983)]{Eggleton1983}
	Eggleton, P.~P.\ 1983, ApJ, 268, 368, \doi{10.1086/160960}

	\bibitem[Eggleton(2012)]{Eggleton2012}
		Eggleton, P.~P. 2012, Journal of Astronomy and Space Sciences, 29, 145, \doi{10.5140/JASS.2012.29.2.145}

	\bibitem[Fabry et~al.(2023)]{Fabry2023}
		Fabry, M., Marchant, P., Langer, N., \& Sana, H. 2023, A\&A, 672, A175, \doi{10.1051/0004-6361/202346277}

	\bibitem[Fabry \& Pr\v{s}a(2025a)]{Fabry2025a}
		Fabry, M., \& Pr\v{s}a, A.\ 2025, ApJ, 994, 7, \doi{10.3847/1538-4357/ae14ed}

	\bibitem[Fabry \& Pr\v{s}a(2025b)]{Fabry2025b}
		Fabry, M., \& Pr\v{s}a, A.\ 2025, ApJ, 995, 19, \doi{10.3847/1538-4357/ae1612}

	\bibitem[Flannery(1976)]{Flannery1976}
		Flannery, B.~P. 1976, ApJ, 205, 217, \doi{10.1086/154266}

	\bibitem[Foreman-Mackey et~al.(2013)]{ForemanMackey2013}
		Foreman-Mackey, D., Hogg, D.~W., Lang, D., \& Goodman, J. 2013, PASP, 125, 306, \doi{10.1086/670067}

\bibitem[{{Gaia Collaboration}(2020)}]{10.5270/esa-1ugzkg7}
{Gaia Collaboration}. 2020, Gaia Early Data Release 3 (Gaia EDR3),  European
  Space Agency, \dodoi{10.5270/esa-1ugzkg7}

	\bibitem[Gaia Collaboration et~al.(2016)]{Gaia2016}
		Gaia Collaboration, Prusti, T., de Bruijne, J. H. J., et~al. 2016, A\&A, 595, A1, \doi{10.1051/0004-6361/201629272}

	\bibitem[Gaia Collaboration et~al.(2021)]{Gaia2021}
		Gaia Collaboration, Brown, A. G. A., Vallenari, A., et~al. 2021, A\&A, 649, A1, \doi{10.1051/0004-6361/202039657}

	\bibitem[Green et~al.(2019)]{Green2019}
		Green, G. M., Schlafly, E., Zucker, C., Speagle, J. S., \& Finkbeiner, D. 2019, ApJ, 887, 93, \doi{10.3847/1538-4357/ab5362}

	\bibitem[Guinan \& Bradstreet(1988)]{GuinanBradstreet1988}
		Guinan, E. F., \& Bradstreet, D. H. 1988, in Dupree, A. K., \& Lago, M. T. V. T., eds, Kinematic Clues to the Origin and Evolution of Low Mass Contact Binaries, p. 345, \doi{10.1007/978-94-009-3037-7\_23}

	\bibitem[Guo et~al.(2022)]{Guo2022}
Guo, D.-F., Li, K., Liu, F., Hu, S.-M., Chen, X., Jiang, Y.-G., Gao, D.-Y., \& Guo, D.-F. 2022, MNRAS, 517, 1928, \doi{10.1093/mnras/stac2811}

	\bibitem[Guo et~al.(2025)]{Guo2025}
Guo, D.-F., Li, K., Liu, F., Wang, L.-H., Li, H.-Z., Chen, X., \& Gao, X. 2025, AJ, 170, 101, \doi{10.3847/1538-3881/ade9c0}

	\bibitem[Horvat et~al.(2018)]{Horvat2018}
		Horvat, M., Conroy, K. E., Pr\v{s}a, A., et~al. 2018, ApJS, 237, 26, \doi{10.3847/1538-4365/aacd0f}

	\bibitem[Hu et~al.(2014)]{Hu2014}
		Hu, S.-M., Han, S.-H., Guo, D.-F., \& Du, J.-J.\ 2014, RAA, 14, 719, \doi{10.1088/1674-4527/14/6/010}
	
	\bibitem[Hurley et al.(2002)]{hurley2002BSE}
		Hurley, J.~R., Tout, C.~A., \& Pols, O.~R. 2002, MNRAS, 329, 897, \doi{10.1046/j.1365-8711.2002.05038.x}

	\bibitem[Hut(1980)]{Hut1980}
		Hut, P. 1980, A\&A, 92, 167

	\bibitem[Jayasinghe et~al.(2018)]{Jayasinghe2018}
		Jayasinghe, T., Kochanek, C. S., Stanek, K. Z., et~al. 2018, MNRAS, 477, 3145, \doi{10.1093/mnras/sty838}

	\bibitem[Jones et~al.(2020)]{Jones2020}
		Jones, D., Conroy, K. E., Sandquist, E. L., et~al. 2020, ApJS, 247, 63, \doi{10.3847/1538-4365/ab7927}

	\bibitem[Kalimeris et~al.(2002)]{Kalimeris2002}
Kalimeris, A., Rovithis-Livaniou, H., \& Rovithis, P. 2002, A\&A, 387, 969, \doi{10.1051/0004-6361:20020456}

	\bibitem[Kuiper(1941)]{Kuiper1941}
		Kuiper, G. P. 1941, ApJ, 93, 133, \doi{10.1086/144252}

	\bibitem[Kwee(1958)]{Kwee1958}
		Kwee, K. K. 1958, BAN, 14, 131

	\bibitem[Kwee \& van Woerden(1956)]{KweeWoerden1956}
		Kwee, K. K., \& van Woerden, H. 1956, BAN, 12, 327
		
	\bibitem[Landau \& Lifshitz(1975)]{Landau1975}
	Landau, L.~D., \& Lifshitz, E.~M.\ 1975, The Classical Theory of Fields (Oxford: Pergamon Press)

	\bibitem[Li et~al.(2004)]{Li2004}
		Li, L., Han, Z., \& Zhang, F. 2004, MNRAS, 355, 1383, \doi{10.1111/j.1365-2966.2004.08457.x}

	\bibitem[Li et~al.(2017)]{Li2017}
Li, K., Hu, S.-M., Chen, X., \& Guo, D.-F.\ 2017, PASJ, 69, 79, \doi{10.1093/pasj/psx064}

	\bibitem[Li et~al.(2020)]{Li2020}
		Li, K., Kim, C.-H., Xia, Q.-Q., et~al. 2020, AJ, 159, 189, \doi{10.3847/1538-3881/ab7cda}

	\bibitem[Li et~al.(2021)]{Li2021}
		Li, K., Xia, Q.-Q., Kim, C.-H., Gao, X., Hu, S.-M., Guo, D.-F., Gao, D.-Y., Chen, X., \& Guo, Y.-N.\ 2021, AJ, 162, 13, \doi{10.3847/1538-3881/abfc53}

	\bibitem[Li et~al.(2022)]{Li2022ELMRCBI}
		Li, K., Gao, X., Liu, X.-Y., et~al. 2022, AJ, 164, 202, \doi{10.3847/1538-3881/ac8ff2}

	\bibitem[Li et~al.(2024)]{Li2024a}
Li, K., Gao, X., Guo, D.-F., Gao, D.-Y., Chen, X., Wang, L.-H., Xin, Y.-X., Han, Y.-X., Kim, C.-H., \& Jeong, M.-J. 2024, A\&A, 692, L4, \doi{10.1051/0004-6361/202451947}

\bibitem[Li et~al.(2024)]{Li2024}
Li, X.-Z., Zhu, Q.-F., Ding, X., Xu, X.-H., et~al.\ 2024, ApJS, 271, 32, \doi{10.3847/1538-4365/ad226a}

	\bibitem[Li et~al.(2025a)]{Li2025a}
		Li, K., Wang, L.-H., \& Gao, X.\ 2025, ApJS, 281, 1, \doi{10.3847/1538-4365/ae064c}

	\bibitem[Li et~al.(2025b)]{Li2025b}
		Li, L.-Z., Li, K., Gao, X., Chen, X.-D., Feng, S., Gao, D.-Y., Guo, D.-F., Chen, X., Gao, X., Sun, G.-Y., Bai, S.\ Y.\ C., \& Esamdin, A.\ 2025, MNRAS, 537, 2258, \doi{10.1093/mnras/staf086}

	\bibitem[Li \& Zhang(2006b)]{LiZhang2006b}
		Li, L., \& Zhang, F. 2006b, MNRAS, 369, 2001, \doi{10.1111/j.1365-2966.2006.10462.x}

	\bibitem[Liu et~al.(2025)]{ELMRCBII}
		Liu, F., Li, K., Gao, X., et~al. 2025, MNRAS, 540, 1290, \doi{10.1093/mnras/staf763}

	\bibitem[Liu \& Yang(2003)]{LiuYang2003}
		Liu, Q.-Y., \& Yang, Y.-L. 2003, ChJAA, 3, 142, \doi{10.1088/1009-9271/3/2/142}

	\bibitem[Lucy(1967)]{Lucy1967}
		Lucy, L. B. 1967, AJ, 72, 813, \doi{10.1086/110452}

	\bibitem[Lucy(1968)]{Lucy1968}
		Lucy, L. B. 1968, ApJ, 151, 1123, \doi{10.1086/149510}

	\bibitem[Mattei \& Saladyga(1999)]{MatteiSaladyga1999}
		Mattei, J. A., \& Saladyga, M. 1999, Observing Variable Stars: A Guide for the Beginner (Cambridge: Cambridge Univ. Press)

	\bibitem[Meng et~al.(2024)]{Meng2024}
		Meng, Z.-B., Wu, P.-R., Yu, Y.-X., Hu, K., \& Xiang, F.-Y. 2024, ApJ, 971, 113, \doi{10.3847/1538-4357/ad571e}

	\bibitem[Montes et~al.(1995)]{Montes1995}
		Montes, D., Fernandez-Figueroa, M. J., de Castro, E., \& Cornide, M. 1995, A\&AS, 114, 287

	\bibitem[Mullan(1975)]{Mullan1975}
		Mullan, D.~J.\ 1975, ApJ, 198, 563, \doi{10.1086/153635}

	\bibitem[O'Connell(1951)]{OConnell1951}
		O'Connell, D. J. K. 1951, PRCO, 2, 85

	\bibitem[Pesta \& Pejcha(2023)]{Pesta2023}
		Pesta, M., \& Pejcha, O.\ 2023, A\&A, 672, A176, \doi{10.1051/0004-6361/202245613}

	\bibitem[Popov \& Petrov(2022)]{Popov2022}
		Popov, V. A., \& Petrov, N. I. 2022, NewA, 97, 101862, \doi{10.1016/j.newast.2022.101862}

	\bibitem[Poro et al.(2024)]{Poro2024a}
		Poro, A., Hedayatjoo, M., Nastaran, M., Nourmohammad, M., et al. 2024, New Astronomy, 110, 102227, \doi{10.1016/j.newast.2024.102227}

	\bibitem[Poro et al.(2026)]{Poro2026}
		Poro, A., Paki, E., Alicavus, F., \& Michel, R. 2026, ApJ, 998, 108, \doi{10.3847/1538-4357/ae3375}

	\bibitem[Pribulla et~al.(2003)]{Pribulla2003}
		Pribulla, T., Kreiner, J. M., \& Tremko, J. 2003, CoSka, 33, 38

	\bibitem[Pr\v{s}a et~al.(2016)]{Prsa2016}
		Pr\v{s}a, A., Conroy, K. E., Horvat, M., et~al. 2016, ApJS, 227, 29, \doi{10.3847/1538-4365/227/2/29}

	\bibitem[Pr\v{s}a \& Zwitter(2005)]{PrsaZwitter2005}
		Pr\v{s}a, A., \& Zwitter, T. 2005, ApJ, 628, 426, \doi{10.1086/430591}

	\bibitem[Qian et~al.(2006)]{Qian2006}
Qian, S.-B., Yang, Y.-G., Zhu, L.-Y., He, J.-J., \& Yuan, J.-Z.\ 2006, Ap\&SS, 304, 25, \doi{10.1007/s10509-006-9114-z}

	\bibitem[Rasio(1995)]{Rasio1995}
		Rasio, F. A. 1995, ApJ, 444, L41, \doi{10.1086/187855}
		
	\bibitem[Reimers(1975)]{Reimers1975}
		Reimers, D.\ 1975, in Problems in Stellar Atmospheres and Envelopes, 229

	\bibitem[Ricker et~al.(2014)]{Ricker2014}
		Ricker, G. R., Winn, J. N., Vanderspek, R., et~al. 2014, Proc. SPIE, 9143, 914320, \doi{10.1117/12.2063489}

	\bibitem[Ruci\'nski(1973)]{Rucinski1973}
		Ruci\'nski, S. M. 1973, AcA, 23, 79

	\bibitem[Rucinski(1993)]{Rucinski1993}
		Rucinski, S. M. 1993, in The Realm of Interacting Binary Stars, ASSL, 177, 111, \doi{10.1007/978-94-011-2416-4\_8}

	\bibitem[Rucinski(2001)]{Rucinski2001}
		Rucinski, S. M. 2001, AJ, 122, 1007, \doi{10.1086/321153}

	\bibitem[Shappee et~al.(2014)]{Shappee2014}
		Shappee, B. J., Prieto, J. L., Grupe, D., et~al. 2014, ApJ, 788, 48, \doi{10.1088/0004-637X/788/1/48}

	\bibitem[Shaw(1994)]{Shaw1994}
		Shaw, J.\ 1994, MmSAI, 65, 95
		
	\bibitem[Sills et~al.(2000)]{Sills2000}
	Sills, A., Pinsonneault, M.~H., \& Terndrup, D.~M.\ 2000, ApJ, 534, 335, \doi{10.1086/308739}
	
\bibitem[{{Skrutskie} {et~al.}(2003){Skrutskie}, {Cutri}, {Stiening},
  {Weinberg}, {Schneider}, {Carpenter}, {Beichman}, {Capps}, {Chester},
  {Elias}, {Huchra}, {Liebert}, {Lonsdale}, {Monet}, {Price}, {Seitzer},
  {Jarrett}, {Kirkpatrick}, {Gizis}, {Howard}, {Evans}, {Fowler}, {Fullmer},
  {Hurt}, {Light}, {Kopan}, {Marsh}, {McCallon}, {Tam}, {Van Dyk}, \&
  {Wheelock}}]{10.26131/irsa2}
{Skrutskie}, M.~F., {Cutri}, R.~M., {Stiening}, R., {et~al.} 2003, 2MASS
  All-Sky Point Source Catalog (PSC),  IPAC, \dodoi{10.26131/IRSA2}

	\bibitem[St\k{e}pie\'n(2006)]{Stepien2006}
		St\k{e}pie\'n, K. 2006, AcA, 56, 347

	\bibitem[St\k{e}pie\'n \& Kiraga(2015)]{StepienKiraga2015}
		St\k{e}pie\'n, K., \& Kiraga, M. 2015, A\&A, 577, A117, \doi{10.1051/0004-6361/201425550}
		
\bibitem[{{SuperWASP Consortium}(2019)}]{10.26133/nea9}
{SuperWASP Consortium}. 2019, SuperWASP Survey Light Curves,  IPAC,
  \dodoi{10.26133/NEA9}

	\bibitem[Terrell \& Wilson(2005)]{TerrellWilson2005}
		Terrell, D., \& Wilson, R. E. 2005, Ap\&SS, 296, 221, \doi{10.1007/s10509-005-4449-4}
		
\bibitem[{{TESS Team}(2022)}]{10.17909/0cp4-2j79}
{TESS Team}. 2022, TESS Calibrated Full Frame Images: All Sectors,  STScI/MAST,
  \dodoi{10.17909/0cp4-2j79}

	\bibitem[Tran et~al.(2013)]{Tran2013}
		Tran, K., Levine, A., Rappaport, S., et~al. 2013, ApJ, 774, 81, \doi{10.1088/0004-637X/774/1/81}

	\bibitem[Wadhwa et~al.(2021)]{Wadhwa2021}
		Wadhwa, S.~S., De Horta, A., Filipovi\'{c}, M.~D., et~al.\ 2021, MNRAS, 501, 229, \doi{10.1093/mnras/staa3637}

	\bibitem[Wadhwa et~al.(2024)]{Wadhwa2024}
		Wadhwa, S.~S., Landin, N.~R., Kosti\'{c}, P., et~al.\ 2024, MNRAS, 527, 1, \doi{10.1093/mnras/stad3129}

	\bibitem[Wilsey \& Beaky(2009)]{Wilsey2009}
Wilsey, N.~J., \& Beaky, M.~M.\ 2009, in Society for Astronomical Sciences 28th Annual Symp., 107

	\bibitem[Yakut \& Eggleton(2005)]{YakutEggleton2005}
		Yakut, K., \& Eggleton, P.~P. 2005, ApJ, 629, 1055, \doi{10.1086/431300}

	\bibitem[Y\i lmaz et~al.(2023)]{Yilmaz2023}
		Y\i lmaz, M., \c{S}enavc\i, H. V., Bahar, E., et~al. 2023, NewA, 101, 102022, \doi{10.1016/j.newast.2023.102022}

	\bibitem[Yang \& Qian(2015)]{YangQian2015}
		Yang, Y.-G., \& Qian, S.-B. 2015, AJ, 150, 69, \doi{10.1088/0004-6256/150/3/69}

	\bibitem[Zhang(2024)]{Zhang2024}
		Zhang, X.-D.\ 2024, Scientific Reports, 14, 13011, \doi{10.1038/s41598-024-63833-y}

	\bibitem[Zhang et~al.(2021)]{Zhang2021}
	Zhang, B., Li, J., Yang, F., Xiong, J.-P., et~al.\ 2021, ApJS, 256, 14, \doi{10.3847/1538-4365/ac0834}

	\bibitem[Zhang et~al.(2026)]{Zhang2026}
Zhang, X., Chen, X., Li, Y., \& Fu, J.\ 2026, ApJ, 998, 55, \doi{10.3847/1538-4357/ae2a2c}

	\bibitem[Zheng et~al.(2021)]{Zheng2021}
		Zheng, S.-Y., Li, K., \& Xia, Q.-Q.\ 2021, MNRAS, 506, 4251, \doi{10.1093/mnras/stab1829}

	\bibitem[Zhou \& Leung(1990)]{ZhouLeung1990}
		Zhou, D.-Q., \& Leung, K.-C. 1990, ApJ, 355, 271, \doi{10.1086/168760}
		
\bibitem[{{ZTF Team}(2025)}]{10.26131/irsa598}
{ZTF Team}. 2025, ZTF Lightcurves,  IPAC, \dodoi{10.26131/IRSA598}

	\bibitem[Zwitter et~al.(2003)]{Zwitter2003}
		Zwitter, T., Munari, U., Marrese, P. M., et~al. 2003, A\&A, 404, 333, \doi{10.1051/0004-6361:20030446}

\end{thebibliography}
\end{document}